\documentclass[11pt]{article}
\pdfoutput=1 
\usepackage{mathtools}
\usepackage{booktabs}
\usepackage[english]{babel}
\usepackage{amsmath,amssymb,amsbsy,amstext, amsthm, simplewick, amsfonts}
\usepackage{graphicx}
\usepackage[small]{caption}
\usepackage{siunitx}
\usepackage{upgreek}
\usepackage{framed}
\usepackage{wrapfig}
\usepackage{multirow}
\usepackage{bbm}
\usepackage[numbers,sort&compress]{natbib}
\usepackage[svgnames,dvipsnames,x11names]{xcolor}
\usepackage[utf8x]{inputenc}
\usepackage{selinput}
\usepackage{bm}
\usepackage{float}
\usepackage{geometry}
\usepackage{yfonts}
\usepackage{caption}
\usepackage{subcaption}
\usepackage{sidecap}
\usepackage{longtable}
\usepackage{anyfontsize}
\usepackage{mathrsfs}

\usepackage{epstopdf}
\usepackage{cancel}
\usepackage{tcolorbox}

\usepackage{tocloft}

\newcommand{\ba}{\begin{eqnarray}}
\newcommand{\ea}{\end{eqnarray}}
\def\xyma{\xymatrix@M.7em}
\def\xymas{\xymatrix@M.1em}

\newcommand{\Comment}[1]{{}}
\definecolor{darkblue}{rgb}{0.15,0.35,0.55}
\definecolor{reddish}{rgb}{0.65, 0.2, 0.2}
\definecolor{darkgreen}{RGB}{50,150,0}
\definecolor{greyish2}{rgb}{.96,.96,.96}
\usepackage[linktocpage=true]{hyperref}
\hypersetup{
colorlinks=true,
citecolor=darkblue,
linkcolor=reddish,
urlcolor=darkblue,
pdfauthor={},
pdftitle={},
pdfsubject={}
}

\DeclareFontFamily{OT1}{rsfs10}{}
\DeclareFontShape{OT1}{rsfs10}{m}{n}{ <-> rsfs10 }{}
\DeclareMathAlphabet{\mathscript}{OT1}{rsfs10}{m}{n}

\def\gsim{ \lower .75ex \hbox{$\sim$} \llap{\raise .27ex \hbox{$>$}} }
\def\lsim{ \lower .75ex \hbox{$\sim$} \llap{\raise .27ex \hbox{$<$}} }
\def\be{\begin{equation}}
\def\ee{\end{equation}}
\def\bea{\begin{eqnarray}}
\def\eea{\end{eqnarray}}

\DeclareMathOperator{\e}{e}

\newcommand{\D}{{\rm d}}

\newcommand{\hypergeom}[2]{
  \mathbin{_{#1}{\sf F}_{#2}} }

\usepackage{latexsym,amsmath,amssymb,epsfig}

\definecolor{greyish}{rgb}{.90,.90,.90}
\definecolor{greyish2}{rgb}{.96,.96,.96}
\usepackage{xcolor,colortbl}
\usepackage{tcolorbox}

\usepackage[all]{xy}

\usepackage{ulem}

\renewcommand\[{\begin{equation}} 
\renewcommand\]{\end{equation}}

\colorlet{shadecolor}{greyish2}

\begin{document}

\vspace{-4cm} 
\begin{flushright}
    MI-HET-830
\end{flushright}
\vspace{0.5cm}

\begin{center}
{\fontsize{21.5}{18} \bf{Ladder Symmetries and Love Numbers of \\[8pt]  Reissner--Nordstr\"om Black Holes}}
\end{center} 

\vspace{.2truecm}

\begin{center}
{\fontsize{13.5}{18}\selectfont
Mudit Rai,${}^{\rm a,b}$  and
Luca Santoni${}^{\rm c}$ 
}
\end{center}
\vspace{.4truecm}

\centerline{{\it ${}^{\rm a}$Pittsburgh Particle Physics, Astrophysics, and Cosmology Center,}}
 \centerline{{\it Department of Physics and Astronomy, University of Pittsburgh, Pittsburgh, USA}} 
 
  \vspace{.3cm}

\centerline{{\it ${}^{\rm b}$Mitchell Institute for Fundamental Physics and Astronomy,}}
 \centerline{{\it Department of Physics and Astronomy, Texas A\&M University, College Station, USA}} 
 
  \vspace{.3cm}
 
 \centerline{{\it ${}^{\rm c}$Universit\'e Paris Cit\'e, CNRS, Astroparticule et Cosmologie,}}
 \centerline{{\it 10 Rue Alice Domon et L\'eonie Duquet, F-75013 Paris, France}}
\vspace{.25cm}

\vspace{.5cm}
\begin{abstract}
\noindent
It is well known that asymptotically flat black holes in general relativity have vanishing tidal Love numbers. In the case of Schwarzschild and Kerr black holes, this property has been shown to be a consequence of a hidden structure of ladder symmetries for the perturbations. In this work, we extend the ladder symmetries to non-rotating charged black holes in general relativity. As opposed to previous works in this context, we adopt a  more general definition of Love numbers, including quadratic operators that mix gravitational and electromagnetic perturbations in the point-particle effective field theory. We show that the calculation of a subset of those couplings in full general relativity is affected by an ambiguity in the split between source and response, which we resolve through an analytic continuation. As a result, we derive a novel master equation that unifies scalar, electromagnetic and gravitational perturbations around Reissner--Nordstr\"om black holes. The equation is hypergeometric and can be obtained from previous formulations via nontrivial field redefinitions, which allow to systematically remove some of the singularities and make the presence of the ladder symmetries more manifest.
\end{abstract}

\newpage

\setcounter{tocdepth}{2}
\tableofcontents
\newpage
\renewcommand*{\thefootnote}{\arabic{footnote}}
\setcounter{footnote}{0}

\section{Introduction}

The direct detection of gravitational waves by the LIGO/Virgo interferometers \cite{LIGOScientific:2016aoc} from merging binary systems has opened a new epoch in the study of gravity and compact objects in the strong-field regime \cite{Bailes:2021tot}. To maximize the science return from the increasing number of gravitational-wave observations and the discovery potential of future detectors~\cite{Saleem:2021iwi,2017arXiv170200786A,Reitze:2019iox,Punturo:2010zz}, ever more accurate waveform templates are necessary~\cite{Sathyaprakash:2019yqt,Maggiore:2019uih,Kalogera:2021bya,Berti:2022wzk}. This requires in particular to have a precise understanding of the conservative and dissipative dynamics of two-body systems~\cite{Buonanno:2022pgc}. In this respect, a relevant role is played by tidal effects~\cite{Flanagan:2007ix}. The tidal deformability of compact objects in a binary system affects the inspiral and in turn the phase of the emitted gravitational-wave signal. Measuring tidal effects will offer important insights on the interior structure of compact objects: it will for instance allow to extract relevant information about the equation of state of neutron stars \cite{Vines:2011ud,Bini:2012gu,Steiner:2014pda,Lattimer:2015nhk,LIGOScientific:2018cki,Iacovelli:2023nbv}, reveal the existence of exotic compact objects~\cite{Cardoso:2017cfl,Franzin:2017mtq,Cardoso:2019rvt,Pani:2019cyc,Chirenti:2020bas,Chia:2023tle}, and potentially disclose unknown aspects of the physics at the horizon of black holes.

The tidal deformation of a self-gravitating body is in general parametrized in terms of complex coefficients. Their real parts are often called Love numbers~\cite{Lovepaper,Will:2016sgx} and describe the conservative induced response of the body. The imaginary parts are instead associated to dissipation. It is well known, as a result of an explicit calculation in general relativity, that the Love numbers of asymptotically flat black holes in four spacetime dimensions are exactly zero. For Schwarzschild black holes, the calculation was originally obtained in~\cite{Fang:2005qq,Damour:2009vw,Binnington:2009bb} (see also \cite{Kol:2011vg,Gurlebeck:2015xpa,Hui:2020xxx,Ivanov:2022qqt}). It was later generalized to the case of Kerr black holes in~\cite{LeTiec:2020spy,LeTiec:2020bos,Chia:2020yla,Charalambous:2021mea}, where it was shown that, in the presence of spin, the tidal response coefficients are purely imaginary. The vanishing of the Love numbers of charged, non-rotating black holes in Einstein--Maxwell theory in four spacetime dimensions was instead pointed out in~\cite{Cardoso:2017cfl,Pereniguez:2021xcj}.

The peculiar property of the vanishing of the induced static response of black holes is even more mysterious when it is formulated in terms of a large-distance effective description for the object \cite{Goldberger:2004jt,Goldberger:2005cd,Goldberger:2009qd,Rothstein:2014sra,Porto:2016pyg,Levi:2018nxp,Goldberger:2020fot,Goldberger:2022ebt,Goldberger:2022rqf}. The idea is based on the simple intuition that, from far away, any object appears in first approximation as a point source. Effects associated to its finite size and its nontrivial internal structure can then be account for in terms of higher-dimensional operators attached to its worldline. In this effective-field-theory (EFT) description of the source, the Love numbers correspond to the coupling constants of higher-dimensional operators. This provides a robust definition of  tidal deformability that is free of ambiguities related to the choice of coordinates and nonlinearities~\cite{Porto:2016pyg,Ivanov:2022hlo,Riva:2023rcm}. On the other hand, the EFT gives a systematic recipe for computing the induced response of the object, based on a standard matching procedure between the EFT and the full solution in general relativity. After the matching, the black hole Love number couplings are found to be zero, appearing therefore to be associated to a naturalness puzzle in the gravitational EFT~\cite{Rothstein:2014sra,Porto:2016zng}. 
Recent investigations suggest that the vanishing of the Love number couplings is perhaps not so puzzling in the end: underlying it, in fact, is a hidden structure of symmetries for the linearized perturbations~\cite{Hui:2021vcv,Hui:2022vbh,BenAchour:2022uqo,katagiri2023vanishing},\footnote{See Refs.~\cite{Charalambous:2021kcz,Charalambous:2022rre} for a different proposal.} which can be used to explain the absence of induced response of black holes in four spacetime dimensions and rule out Love number couplings in the EFT \cite{Hui:2021vcv}. 

The goal of the present work is to extend the ladder symmetries of Schwarzschild and Kerr spacetimes  to Reissner--Nordstr\"om  black holes, and show that they can be analogously used to rule out Love number couplings. 
We start off by clarifying some previously unnoticed properties of the equations describing the dynamics of the gravito-electromagnetic perturbations.
We will adopt the Chandrasekhar formulation, where the equations of the gravitational and electromagnetic perturbations are decoupled and take, in the static limit, a Fuchsian form with four and six regular singular points for the axial and polar sector, respectively. We will show that all but three singular points of the Fuchsian differential equations are removable in four spacetime dimensions, and we will provide the explicit expressions of the field redefinitions that allow to recast the equations in the standard hypergeometric form. 
We will then solve them analytically in terms of finite polynomials and show in full generality that the Love numbers vanish.\footnote{Our procedure provides an alternative and more systematic explanation of why the solution in \cite{Pereniguez:2021xcj} in $D=4$, where $D$ is the number of spacetime dimensions, was found to be in the form of a finite polynomial, despite the fact that the equations in the vector sector are of the Heun type.} Our results are  in agreement with the ones in \cite{Cardoso:2017cfl,Pereniguez:2021xcj}, although, in contrast to those references, we adopt here a wider definition of tidal response coefficients, based on the point-particle EFT (see eq.~\eqref{eq:Sfinitesize}), which includes also operators that mix gravitational and electromagnetic perturbations. Related to this,  in  the derivation of the static solutions, we will point out a subtlety related to an ambiguity in the  source/response split, which we will address via an analytic continuation, and explicitly perform the matching with the point-particle EFT.   
Finally, we will show that the symmetries of the hypergeometric equation act on the static Reissner--Nordstr\"om perturbations as  ladder symmetry generators, which can be used to understand why the static solutions take the form of finite polynomials and why the induced response vanishes.
Our result  generalizes the one of \cite{Hui:2021vcv} to black holes with charge, and the one of \cite{Berens:2022ebl} to spin-1 and spin-2 fields.

\vspace{-.5cm}
\paragraph{Outline:} The paper is organized as follows. In section~\ref{sec:axial} we briefly review the static equations for the axial perturbations of Reissner--Nordstr\"om black holes. In section~\ref{sec:masterZ} we  explicitly show that one of the singularities of the radial differential equations describing the axial sector can be removed, and we derive a general master hypergeometric equation for the Reissner--Nordstr\"om static perturbations. In section~\ref{sec:analytic} we discuss a potential ambiguity in the definition of the Love numbers, which we resolve by performing an analytic continuation that allows us to unambiguously disentangle the response from the subleading tail of the external tidal source. In sections~\ref{sec:polardual} and \ref{sec:scalarfield} we extend the results to the polar sector and to spin-0 perturbations, respectively, showing that their dynamics is also described by the same master equation introduced in section~\ref{sec:masterZ}. The ladder symmetries are then derived in section~\ref{sec:laddersec}. Finally, we discuss the matching with the point-particle effective field theory in section~\ref{sec:matching}. Some technical aspects of the calculations are collected in appendices~\ref{app:perturbations} and \ref{app:masterequation}.

\vspace{-.5cm}
\paragraph{Conventions:} Throughout we use the mostly-plus metric signature, and work in natural units $\hbar=c=1$. We also set $G=1$.
We denote $4$-dimensional spacetime indices using Greek letters, e.g., $\mu,\nu, \rho,\cdots$, denote $3$-dimensional spatial indices by Latin letters from the beginning of the alphabet, e.g., $a,b,c,\cdots$, and we denote angular indices on the $2$-sphere using Latin indices from the middle of the alphabet, e.g., $i,j,k,\cdots$.

\section{Perturbations of Reissner--Nordstr\"om black holes: axial sector}
\label{sec:axial}

Reissner--Nordstr\"om black holes are static solutions to the Einstein--Maxwell equations~\cite{Chandrasekhar:1985kt}. They are described by the following  background 
metric $\bar{g}_{\mu\nu}$,
\begin{equation}
\D s^2 = \bar{g}_{\mu\nu}\D x^\mu \D x^\mu  \equiv - \frac{\Delta(r)}{r^2} \D t^2 + \frac{r^2}{\Delta(r)}\D r^2 + r^2 \left( \D \theta^2 + \sin^2\theta \D \varphi^2 \right) \, ,
\label{gmunuRN}
\end{equation}
with 
\begin{equation}
\Delta(r) \equiv  r^{2}-2Mr+Q^{2} \, ,
\end{equation}
where $M$ and $Q$ denote the mass and the charge of the black hole respectively, while the background Maxwell 4-potential is
\begin{equation}
\bar A_\mu = \left( \frac{Q}{r}, 0,0,0 \right) \,.
\label{Amu}
\end{equation}
The event and Cauchy horizons of the black hole are determined by the roots of $\Delta(r)=0$. Denoting them with $r_+ $ and $r_-$ respectively, they  are given by the following expressions in terms of the black hole mass and charge:
\begin{equation}
r_\pm = M \pm \sqrt{M^2 -Q^2}\, .
\end{equation}

The dynamics of the fluctuations can be studied by perturbing,  in the Einstein--Maxwell equations, the metric tensor $g_{\mu\nu}$  and the vector potential $A_\mu$ around the background solutions $\bar{g}_{\mu\nu}$ and $\bar{A}_\mu$~\cite{Zerilli:1974ai,Moncrief:1974gw,Moncrief:1974ng,Moncrief:1975sb}.  
The spherical symmetry of the background  ensures that the equations of motion of the axial and polar components of the perturbations decouple at linear order. For this reason they can be studied separately. We will start in this section by focusing on the axial components, while we will discuss the polar sector in section~\ref{sec:polardual}.

Using the notation of~\cite{Chandrasekhar:1985kt},  the radial  equations for the axial sector can be cast, in the static limit, in the form\footnote{The notation in eqs.~\eqref{eq:ChandraH1H2}, as well as in eqs.~\eqref{fr12} and \eqref{eq:Heun} below,  mirrors the one of Ref.~\cite{Chandrasekhar:1985kt} except for the fact that, for simplicity, we drop a superscript $-$ to denote the axial fields $H_i$ and $Z_i$. We will reintroduce it when needed below.}
\begin{subequations}
\label{eq:ChandraH1H2}
\begin{align}
 r^2\frac{\D}{\D r}\left[ \frac{\Delta(r)}{r^2}  H_{2}'(r)\right]  -\frac{2}{r}\left[ \left(\frac{r\,\eta(\ell)}{2}-3M+\frac{2Q^{2}}{r}\right) H_{2}(r) +Q\mu(\ell)H_{1}(r)\right] & =0 \, , 
\label{eq:ChandraH1H21}
 \\ 
 r^2\frac{\D}{\D r}\left[ \frac{\Delta(r)}{r^2}  H_{1}'(r)\right] -\frac{2}{r}\left[ \left(\frac{r\,\eta(\ell)}{2}+\frac{2Q^{2}}{r}\right)H_{1}(r) +Q\mu(\ell)H_{2}(r)\right] & =0 \, ,
\label{eq:ChandraH1H22}
\end{align}
\end{subequations}
where $\eta(\ell)\equiv \ell(\ell+1)$ and $\mu(\ell)\equiv \sqrt{\eta(\ell)-2}$, and where $H_1$ and $H_2$ are related to the perturbations of the metric tensor, $\delta g_{\mu\nu}= g_{\mu\nu}-\bar{g}_{\mu\nu}$, and of the electromagnetic vector potential, $\delta A_\mu = A_\mu-\bar{A}_\mu$. We refer to \cite{Chandrasekhar:1985kt} for the precise definitions of $H_1$ and $H_2$ (see also appendix~\ref{app:perturbations} for a short summary of the relation between $H_1$ and $H_2$ and the physical perturbations $\delta g_{\mu\nu}$ and $\delta A_\mu$).
It is convenient to introduce the $\ell$-dependent quantities 
\begin{subequations}
\begin{align}
q_{1}&=3M+\sqrt{9M^{2}+4Q^{2}\mu^{2}(\ell)} \, , 
\\
q_{2} & =6M-q_{1}=3M-\sqrt{9M^{2}+4Q^{2}\mu^{2}(\ell)} \, ,
\end{align}
\label{eq:q1q2}
\end{subequations}
which satisfy $q_{1}q_{2}=-4Q^{2}\mu^{2}(\ell)$.
Using the following field redefinition,\footnote{Note that the redefinitions \eqref{fr12} slightly differ with respect to the ones introduced in \cite{Chandrasekhar:1985kt}. The expressions of \cite{Chandrasekhar:1985kt} can be recovered by  rescaling the field variable $Z_2$ in eqs.~\eqref{fr12} as $Z_2 \rightarrow \frac{2Q\mu}{q_1}Z_2$.}
\begin{subequations}
\label{fr12}
\begin{align}
     H_1(r) & =\frac{ Z_1(r)- Z_2(r)}{q_1- q_2} \, ,
     \label{fr1}
    \\ 
    H_2(r) & = \frac{1}{2Q\mu} \frac{q_1 Z_2(r) - q_2 Z_1(r)}{q_1- q_2}\, ,
    \label{fr2}
\end{align}
\end{subequations}
eqs.~\eqref{eq:ChandraH1H2} decouple,  and boil down to~\cite{Moncrief:1974gw,Chandrasekhar:1985kt}
\begin{equation}
r^2\frac{\D}{\D r}\left[ \frac{\Delta}{r^2}  Z_{i}'(r)\right] -\frac{1}{r} \left(r\eta(\ell)+\frac{4Q^{2}}{r}-q_{j}\right) Z_i(r)=0 \, , \qquad(i,j\in\{1,2\}, \, i\neq j) \, .
\label{eq:Heun}
\end{equation}
Eqs.~\eqref{eq:Heun} are second-order, ordinary differential equations with four regular singular points at the locations  $\{0,r_{-},r_{+},\infty\}$. Hence, they belong to the Heun class~\cite{slavjanov2000special,MR1392976}. The presence of four singularities makes the equations less amenable to be studied analytically, with respect for instance to the Schwarzschild (i.e., $Q=0$) case, where they take  instead a hypergeometric form in the static limit. Connection formulas relating Frobenius solutions at different singular points---which are necessary to compute the induced static  response---are more involved to obtain, and can often be written analytically in closed form only perturbatively in some limits (see, e.g., \cite{Bonelli:2021uvf,Bonelli:2022ten,Lisovyy_2022}).  Although these aspects are well known in the literature, the new twist that we will add here is that one of the singularities of  the Heun equations \eqref{eq:Heun} can actually be removed,  reducing \eqref{eq:Heun} to a hypergeometric equations. This will significantly simplify the analysis of the static perturbations, allowing us to derive general analytic solutions, as we shall now show explicitly.

\section{Master hypergeometric equation for Reissner--Nordstr\"om black hole perturbations}
\label{sec:masterZ}

Fuchsian equations are linear differential equations with  coefficients that are holomorphic functions of the variable except for a finite number of  singular points. 
The Heun equation is an example, having four regular singularities and corresponding therefore to the next case in the Fuchsian series after the hypergeometric equation~\cite{MR1392976}.
With their confluent subcases, Fuchsian equations are found to describe  a wide variety of problems in physics. 

There are special cases in which, in a given singular point of a  Fuchsian equation, both linearly independent solutions happen to be holomorphic. In these cases, the singularity can be removed via a field redefiniton and, for this reason, it is often called `apparent'~\cite{slavianov2000special}. Removing a singular point can be particularly convenient, as it allows one to more easily find the solutions of the equation and study their properties. One notorious example is given by the Zerilli equation for parity-even perturbations around a Schwarzschild black hole~\cite{Zerilli:1970se}. In the static limit the equation is of the Heun type, but one of the four singularities is apparent. As a result, the equation can be recast in a hypergeometric form~\cite{1971ApJ...166..175I,1971ApJ...166..197F,Hui:2020xxx}. In four spacetime dimensions,  the field redefinition that  removes the apparent singularity is related to the Chandrasekhar transformation~\cite{10.2307/78902,Chandrasekhar:1985kt}: the latter maps the Zerilli equation into the Regge--Wheeler equation for the odd perturbations~\cite{Regge:1957td}, which is  in fact hypergeometric in the static limit and can therefore be easily solved analytically. 

The goal of this section is to show that something similar happens  with Reissner--Nordstr\"om black holes. Although both the even and odd  static equations of Reissner--Nordstr\"om perturbations have, at first sight, more than  three singularities, they  can   all in fact  be reduced to a single master hypergeometric equation with three regular singular points only. We will start here from the axial sector described by eqs.~\eqref{eq:Heun}.

It is convenient to first perform the following change of variable,
\begin{equation}
z \equiv \frac{r-r_{-}}{r_{+}-r_{-}}=\frac{r-\left(M-\sqrt{M^{2}-Q^{2}}\right)}{2\sqrt{M^{2}-Q^{2}}} \, ,
\label{eq:defzz}
\end{equation}
and introduce
\begin{equation}
2\mathbb{\mathcal{R}}+1  \equiv \frac{M}{\sqrt{M^{2}-Q^{2}}} \, ,
\end{equation}
where $\mathbb{\mathcal{R}}$ is a non-negative, real number.
As a result, the singularities are shifted from  $\{0,r_{-},r_{+},\infty\}$ to $\{-\mathcal{R},0,1,\infty\}$ in the new variable $z$. 
Eqs.~\eqref{eq:Heun} now take the form
\begin{equation}
\partial_z^2 Z_i +\frac{(2z\mathcal{R}+z-\mathcal{R})}{z(z-1)(z+\mathcal{R})} \partial_z Z_i + \frac{q_j (2 \mathcal{R}+1) (z+\mathcal{R})-2 M \left(\eta z^2+2 \mathcal{R} (\eta z+2)+(\eta+4) \mathcal{R}^2\right)}{2 M (z-1) z (z+\mathcal{R})^2} Z_i  =0 \, ,
\label{eq:normalpsi}
\end{equation}
where $i,j=1,2$ and $j\neq i$.

Following, e.g., Ref.~\cite{eremenko2018fuchsian}, it is straightforward to check that the point $z=-\mathcal{R}$ satisfies all the conditions   that are sufficient for it to be a removable singularity of the Heun equation \eqref{eq:normalpsi}.   We 
leave the details of this check to appendix~\ref{app:masterequation}. Once it is established that a singularity is apparent, it is not hard to find the field redefinition that removes it. In the case of eq.~\eqref{eq:normalpsi}, such field redefinition takes the form
\begin{equation}
Z_i(r(z))=(z+\mathcal{R})^{-1}\left[\left(z\frac{\D}{\D z}\right)^{4}+A_i\left(z\frac{\D}{\D z}\right)^{3}+B_i\left(z\frac{\D}{\D z}\right)^{2}+C_i\left(z\frac{\D}{\D z}\right)+D\right]F(z) \, ,
\label{eq.yhyp:main}
\end{equation}
where $A_i$, $B_i$, $C_i$ and $D$ are constant coefficients. Choosing them as follows (see also appendix~\ref{app:masterequation}),
\begin{subequations}
\label{eq:ABCDi:main}
\begin{align}
     A_i & = -\frac{3 (q_i+2 q_j (\mathcal{R}+1))}{(\mathcal{R}+1) (q_i+ q_j)} \, , \\ 
    B_i & = \frac{q_i (2-7 \mathcal{R})+11 q_j (\mathcal{R}+1)}{(\mathcal{R}+1) (q_i+q_j)} \, , \\ 
    C_i & = -\frac{3 \left(q_i \mathcal{R} (\eta+2 (\eta-1) \mathcal{R}-2)+2 q_j (\mathcal{R}+1)^2\right)}{(\mathcal{R}+1)^2 (q_i+q_j)} \, , \\ 
    D  & = -\frac{ \eta\,\mu^2\, \mathcal{R}^2}{(\mathcal{R}+1)^2} \, ,
\end{align}
\end{subequations}
with $i,j=1,2$ and $j\neq i$, it is easy to check that the Heun equation \eqref{eq:normalpsi} 
 is equivalent to  the following simple hypergeometric equation:\footnote{Note that all the information about the removed $z=-\mathcal{R}$ ($r=0$) singularity remains fully encoded in the overall prefactor $(z+\mathcal{R})^{-1}$ of the field redefiniton \eqref{eq.yhyp:main}.}
\begin{shaded}
\begin{equation}
z(1-z)F''(z)+(1+2z)F'(z)+[\ell(\ell+1)-2]F(z)=0 \, .
\label{eq:HyperG:main}
\end{equation}
\end{shaded}
We stress that there is no index $i$ on $F$, nor in the parameters of \eqref{eq:HyperG:main}: in other words, the axial perturbations $Z_1(r)$ and $Z_2(r)$ satisfy the \textit{same} master hypergeometric equation \eqref{eq:HyperG:main}. Dependence on $q_1$ and $q_2$ is only through the coefficients \eqref{eq:ABCDi:main} of the field redefinition \eqref{eq.yhyp:main}, which links \eqref{eq:HyperG:main} to \eqref{eq:normalpsi}. 

Note that the transformation \eqref{eq.yhyp:main} can be greatly simplified by getting rid of the higher derivatives on $F$ using its equation \eqref{eq:HyperG:main}. After straightforward manipulations, one finds
\begin{multline}
    Z_i(r(z))= \frac{z F'(z)}{z+\mathcal{R}}  \left[z\frac{ A_i (1-z) (\eta -(\eta +7) z+1)+3 B_i (z-1)^2+2 \eta +(6 \eta +15 )z^2-8 ( \eta  +1) z-1}{(z-1)^3}+C_i\right]  
\\
    + \frac{F(z)}{z+\mathcal{R}} \left[ z (\eta -2) \frac{ A_i \left(3 z^2-4 z+1\right)+B_i (z-1)^2+(\eta +7) z^2-(\eta  +6) z+1}{(z-1)^3}+D\right] \, ,
\end{multline}
which is in the form of a  generalized Darboux transformation, e.g.~\cite{Glampedakis:2017rar}. 

The master equation \eqref{eq:HyperG:main} can be easily solved in terms of hypergeometric functions. Since $\ell$ is an integer, the equation is degenerate and the most general solution of \eqref{eq:HyperG:main} is given by the following superposition of independent  hypergeometric functions~\cite{bateman1953higher}:
\begin{equation}
F(z)= c_{\ell} (z-1)^{4}\hypergeom{2}{1}(2-\ell,\ell+3,1;z) + d_{\ell}\, (-z)^{1-\ell}  \hypergeom{2}{1}(\ell-1,\ell-1, 2\ell+2;z^{-1})  \, ,
\label{eq:sol2F1}
\end{equation}
where $c_{\ell}$ and $d_{\ell}$ are arbitrary integration constants. Note that, when \eqref{eq:sol2F1} is used to write the solution for  $Z_i$,  there will be  four  constants in total---which we will denote below with $c_{i,\ell}$ and $d_{i,\ell}$ with $i=1,2$---two for each  $Z_i$.

Using eqs.~\eqref{fr12} and \eqref{eq:CardosotoChandra}, it is then  straightforward to write the solutions for the gravitational and electromagnetic potentials $h_0$ and $u_4$ (see eqs.~\eqref{eq:hmunupert_GR} and \eqref{eq:EMpert_GR} for their definitions) in full generality, for arbitrary $\ell$, as
\begin{align}
h_0(r) & = \frac{2Q \Delta(r)}{\eta(\ell)\mu(\ell) r^2} (r H_2(r))' =\frac{\Delta(r)}{\eta(\ell)\mu(\ell)^2 r^2 }\frac{ \left[r(q_1 Z_2(r)-q_2 Z_1(r))\right]'  }{ q_1-q_2 } \, , \label{eqh0m1} \\ 
u_4(r) & = Q H_1(r) = Q\frac{  Z_1(r)- Z_2(r)}{q_1-q_2} \, ,
\label{equ4m1}
\end{align}
where the solutions for $Z_1(r)$ and $Z_2(r)$ are given by eqs.~\eqref{eq.yhyp:main} and \eqref{eq:sol2F1}.

We are interested  in  the static response. Therefore, we are going to fix the integration constants $c_{i,\ell}$ and $d_{i,\ell}$ of the boundary-value problem in such a way that the solution \eqref{eq:sol2F1} asymptotes a static tidal field at large distances and  is regular at the black hole horizon~\cite{Fang:2005qq,Damour:2009vw,Binnington:2009bb}. One can easily check that this amounts to setting $d_{i,\ell}=0$, while $c_{i,\ell}$ correspond to the tidal field amplitudes~\cite{Cardoso:2017cfl}.
Using the series representation of the hypergeometric function, we find\footnote{For an alternative way of writing the solution for $Z_i(r)$, see appendix~\ref{app:masterequation}.}
\begin{multline}
Z_i(r)= \frac{c_{i,\ell}}{r}  
\left[\left((r-r_-)\frac{\D}{\D r}\right)^{4}+A_i\left((r-r_-)\frac{\D}{\D r}\right)^{3}+B_i\left((r-r_-)\frac{\D}{\D r}\right)^{2}+C_i\left((r-r_-)\frac{\D}{\D r}\right)+D\right] 
\\
\times (r-r_+)^4\sum_{n=0}^{\ell-2}(-1)^n \frac{(\ell-2)!}{n!(\ell-n-2)!}  \frac{\Gamma(\ell+n+3)}{\Gamma(n+1)\Gamma(\ell+3)} 
\left(  \frac{r-r_{-}}{r_{+}-r_{-}} \right)^n \, ,
\label{eq:Zihyper}
\end{multline}
which is manifestly a finite polynomial in $r$ with only positive powers, except for a $1/r$ term, for each $i=1,2$.
At large distances, $Z_i(r)$ scale as $\sim r^{\ell+1}$, correctly reproducing the tidal field profile on flat space.
After plugging into the gravitational and electromagnetic potentials  \eqref{eqh0m1} and \eqref{equ4m1}, one can extract the static solutions for $h_0$ and $u_4$, and the induced odd response of the Reissner--Nordstr\"om black hole. 
For instance, as an illustrative example, let us fix $\ell=2$. Using the explicit solution \eqref{eq:Zihyper}, the potentials \eqref{eqh0m1} and \eqref{equ4m1} take the form
\begin{align}
\label{eq:h0l2}
h_0^{\ell=2}(r) & = {\cal{A}}_{h_0} r^3
-   \left(2 {\cal{A}}_{h_0} M + \frac{1}{2} {\cal{A}}_{u_4} Q\right) r^2
+ Q\left( {\cal{A}}_{h_0} Q+ {\cal{A}}_{u_4} M \right) r
- \frac{ {\cal{A}}_{u_4} Q^3 }{3}\left(  1
+\frac{ M }{r}
-\frac{ Q^2}{2 r^2} \right) ,
\\
\label{eq:u4l2}
u_4^{\ell=2}(r)
& = {\cal{A}}_{u_4} r^3
- \frac{3}{2}  ( {\cal{A}}_{h_0} Q + {\cal{A}}_{u_4} M) r^2
+ \frac{3Q^2}{2} \left({\cal{A}}_{h_0} Q+ {\cal{A}}_{u_4} M \right)-\frac{{\cal{A}}_{u_4} Q^4}{r} \,  ,
\end{align}
where we conveniently redefined the arbitrary  integration constants $c_{i,\ell=2}$ into ${\cal{A}}_{h_0}$ and ${\cal{A}}_{u_4}$, corresponding to the amplitudes of the gravitational and electromagnetic external fields, respectively.\footnote{Assuming that the external source is a purely gravitational quadrupolar tidal field, we shall set  ${\cal{A}}_{u_4}=0$ in  the solutions \eqref{eq:h0l2} and \eqref{eq:u4l2}, which reduce to  simply $ h_0^{\ell=2}(r)={\cal{A}}_{h_0}  r \Delta(r)$ and $u_4^{\ell=2}(r) = - \frac{3}{2}{\cal{A}}_{h_0}  Q (r^2-Q^2)$, in agreement with~\cite{Cardoso:2017cfl}.}
Note that, as opposed to the Schwarzschild case ($Q=0$) where the solutions are  polynomials with no inverse powers of $r$, the presence of a nonzero charge is responsible here for a decaying falloff, which is proportional to $Q$ and is  induced by the  mixing between the gravitational and electromagnetic perturbations on the curved Reissner--Nordstr\"om background. The falloff is at most $1/r^2 $ for all $\ell$ (see eqs.~\eqref{eqh0m1}, \eqref{equ4m1} and \eqref{eq:Zihyper}): this implies that, by simple power counting, the static response coefficients  vanish for $\ell\geq3$, as we will more precisely assess by explicitly performing the matching with the point-particle EFT in section~\ref{sec:matching}. The case $\ell=2$, written explicitly in eqs.~\eqref{eq:h0l2} and \eqref{eq:u4l2}, requires  additional care. In fact, the $1/r^2 $ falloff in \eqref{eq:h0l2} resembles at face value an induced quadrupole in the gravitational sector, generated by an external magnetic field with amplitude ${\cal{A}}_{u_4}$. (In the language of the EFT, this would correspond to a non-vanishing coupling of the  quadratic $\ell=2$ operator the mixes the magnetic field  and the odd component of the Weyl tensor, see eq.~\eqref{eq:Sfinitesize} below.) In the following section, we show that this is not true: by carefully performing an analytic continuation in the solution, we will in fact conclude that such a falloff actually corresponds to a subleading correction to the external source on the curved background, and should not be interpreted as an induced response.

\section{Analytic continuation and the Reissner--Nordstr\"om Love numbers}
\label{sec:analytic}

To answer the question of whether the $1/r^2 $ falloff in \eqref{eq:h0l2} corresponds to a nonzero quadrupolar tidal deformation, we will rederive the static solutions for the perturbations by performing an analytic continuation in $\ell\in \mathbb{R}$. The analytic continuation will allow us to disentangle the tidal source from the response field, and provide a clear interpretation for the decaying terms in \eqref{eq:h0l2} and \eqref{eq:u4l2}. 
Note that a similar procedure has been employed before in~\cite{LeTiec:2020spy,LeTiec:2020bos,Charalambous:2021mea} to extract the dissipative response of Kerr black holes in general relativity. To the best of our knowledge, this subtlety for Reissner--Nordstr\"om black holes was not addressed before in the literature. 

Let us start again from our master equation \eqref{eq:HyperG:main}. Instead of solving it for integer $\ell$ as in \eqref{eq:sol2F1}, we will now perform an analytic continuation in $\ell$ to real (non-integer)  values.\footnote{We stress that the derivation of the master equation \eqref{eq:HyperG:main}, and in particular the procedure of removing the singularity from the original Heun equation, did not depend at any point on the fact that $\ell $ was integer or not.} In the analytic continuation sense, the hypergeometric equation is non-degenerate and the independent solutions are given by the fundamental ones~\cite{Bateman:100233}: 
\begin{equation}
F(z)= (z-1)^4 z^{-\ell-3} \left[ \tilde c_\ell \, \hypergeom{2}{1}\left(\ell+3,\ell+3,2 \ell+2;z^{-1}\right)-\tilde d_\ell \,  z^{2 \ell+1}  \hypergeom{2}{1}\left(2-\ell,2-\ell,-2 \ell;z^{-1}\right)\right] \, ,
\label{Fanal}
\end{equation}
where $\ell\in \mathbb{R}$, and where $\tilde c_\ell$ and $\tilde d_\ell$ are arbitrary integration constants. Regularity at the horizon, $z=1$, fixes
\begin{equation}
\tilde d_\ell = \tilde c_\ell \frac{ \Gamma (2-\ell)^2 \Gamma (2 \ell+2)}{\Gamma (-2 \ell) \Gamma (\ell+3)^2} \, .
\end{equation}
Plugging it back into \eqref{Fanal}, the physical solution for $F(z)$ takes the form
\begin{multline}
F(z) = {c}_\ell (z-1)^4 z^{-\ell-3} \bigg[ z^{2 \ell+1} \hypergeom{2}{1}\left(2-\ell,2-\ell,-2 \ell;z^{-1}\right)
\\
- \frac{\Gamma (-2 \ell) \Gamma (\ell+3)^2 }{\Gamma (2-\ell)^2 \Gamma (2 \ell+2)} \hypergeom{2}{1}\left(\ell+3,\ell+3,2 \ell+2;z^{-1}\right) \bigg]
\label{Fanal2}
\end{multline}
where we rescaled the overall amplitudes, $\tilde{c}_\ell \mapsto {c}_\ell$, for practical convenience. Note that, taking the limit of integer $\ell$ in \eqref{Fanal2}, one correctly recovers the result \eqref{eq:Zihyper} obtained in section~\ref{sec:masterZ}. For instance, for $\ell\rightarrow2$, eq.~\eqref{Fanal2} boils down to $F_{\ell=2}(z)={c}_\ell (z-1)^4$, in agreement with eq.~\eqref{eq:sol2F1} with $d_\ell=0$. The two approaches give consistent results at the level of the solution $F(z)$; the crucial difference however is that, as we shall see explicitly, the analytic continuation allows to correctly interpret the falloffs, and unambiguously distinguish source and response, which might not be so obvious from the expressions obtained before assuming integer $\ell$.

From eq.~\eqref{Fanal2}, we can obtain the $Z_i$ fields in \eqref{eq.yhyp:main} and in turn the odd gravitational and magnetic potentials $h_0$ and $u_4$ in eqs.~\eqref{eqh0m1} and \eqref{equ4m1}. 
Let us focus on $h_0$ only, which is the one that for $\ell=2$  contains the $1/r^2$ falloff (see eq.~\eqref{eq:h0l2}), which one could be potentially tempted to interpret as an induced quadrupolar  tidal deformation.
After plugging the explicit solution \eqref{Fanal2} for $F$, the solution for the gravitational potential $h_0$ takes the following asymptotic expansion at large distances, up to an overall irrelevant constant factor:
\begin{equation}
h_0(r\rightarrow\infty) \sim  \frac{  (\ell+2)^2 \mu(\ell)^{-2}}{2 (\mathcal{R}+1)^2(q_1-q_2)} 
\bigg[ r^{ \ell+1} \left( 1+ \dots \right)
+\frac{\Gamma (1-2 \ell) \Gamma (\ell+3)^2 \rho_\ell}{2 \ell (\ell+2)^2\Gamma (1-\ell)^2 \Gamma (2 \ell+2)}  r^{-\ell} \left( 1+ \dots \right) \bigg],
\label{h0largeran}
\end{equation}
where dots denote subleading polynomial  corrections of the form $1/r^n$ (where $n$ is a positive integer) to the corresponding leading branch of solution, and where we defined for convenience the constant
\begin{equation}
\rho_{\ell} \equiv 1+ \frac{2 (2 \ell+1) (2 \mathcal{R}+1) (c_{1,\ell} q_2-c_{2,\ell} q_1)}{(c_{2,\ell} q_1 - c_{1,\ell} q_2)\left[\ell (4 \mathcal{R}+2)+2 \mathcal{R}+1\right] - (c_{2,\ell} q_1 + c_{1,\ell} q_2) \sqrt{4 (2 \ell+1)^2 \mathcal{R} (\mathcal{R}+1)+9}} \, ,
\end{equation}
where the second term on the right-hand side should be thought of as the ratio between the external magnetic field amplitude and the amplitude of the external gravitational tidal potential $h_0$. 
In \eqref{h0largeran}, the piece $r^{ \ell+1} \left( 1+ \dots \right)$ corresponds to the series expansion of the external tidal field, while the response is encoded in $r^{-\ell} \left( 1+ \dots \right)$. 
Note that, if $\ell$ is a positive  real, non-integer number, it is clear from \eqref{h0largeran} that the subleading terms to the tidal source $r^{\ell+1}$ will never overlap with the second series of terms in \eqref{h0largeran}; in particular, they will not generate any $r^{-\ell}$ falloff. An overlap is possible only  when $\ell$ is integer: in this case, there is a mixing  between the $r^{-\ell}$ series and the tail of the external tidal field, introducing  a degeneracy at the level of the source/response split and making therefore the  identification of the Love numbers potentially ambiguous. Such an ambiguity does not arise however for real $\ell$, in which case we can confidently identify the induced multipole moments as the coefficients
\begin{equation}
\lambda_{\ell\in \mathbb{R}} = \frac{\Gamma (1-2 \ell) \Gamma (\ell+3)^2 \rho_\ell}{2 \ell (\ell+2)^2\Gamma (1-\ell)^2 \Gamma (2 \ell+2)} \, .
\label{eq:lambdaREAL}
\end{equation}
Once the tidal response coefficients  are extracted for real $\ell$, we can then safely take the limit of integer $\ell$ in $\lambda_{\ell\in \mathbb{R}}$ and compute the physical Love numbers, which are found to vanish for all $\ell\in\mathbb{N}$, $\ell\geq2$:
\begin{equation}
\lambda_{\ell\in \mathbb{N}} = 0 \, .
\end{equation}
This shows clearly that the $1/r^2$ falloff in the solution \eqref{eq:h0l2} for $\ell=2$ does not correspond to a gravitational quadrupole induced by an external magnetic field via the $\delta g_{\mu\nu}$ -- $A_\mu$ mixing, but it rather belongs to the subleading tail of the source.

\section{Polar sector: Chandrasekhar duality and equality of Love numbers }
\label{sec:polardual}

The conclusions of the previous sections---i.e.,  reduction of the perturbation equations to the master equation \eqref{eq:HyperG:main}, static solutions and analytic continuation---can be easily generalized to the polar sector. Instead of proceeding as prescribed in appendix~\ref{app:masterequation}, a more convenient strategy is however to take advantage of the Chandrasekhar duality~\cite{Chandrasekhar:1985kt}. Such duality provides a mapping between the axial and polar equations, relating even solutions to odd ones, and viceversa. 

The static equations for the radial polar fields are (see appendix~\ref{app:perturbations} for definitions and, e.g., \cite{Cardoso:2017cfl} for a derivation\footnote{We take the opportunity to correct a typo in the  equation for $u_1$ in Ref.~\cite{Cardoso:2017cfl}.})
\begin{subequations}
\begin{align}
\frac{\D}{\D r}\left[ \Delta(r) H_{0}'(r) \right] +\left[\frac{4(Q^{2}-M^{2})}{\Delta(r)}-\frac{2Q^{2}}{r^{2}} -\eta(\ell) \right] H_{0}(r)
{-}\frac{4Q}{r}\left[u'_{1}(r)+\frac{\left(Q^{2}-r^{2}\right)u_{1}(r)}{r\,\Delta(r)}\right] &  =0 \, ,\\
 u_{1}''(r)+\frac{\frac{4Q^{2}}{r^{2}}-\eta(\ell)}{\Delta(r)}u_{1}(r){-}\frac{Q}{r}\left[H'_{0}(r)-\frac{2\left(Q^{2}-Mr\right)}{r\Delta(r)}H_{0}(r)\right] & =0 \, .
\end{align}
\label{eq:polareqs}
\end{subequations}

As in the case of Schwarzschild black holes, the linearized equations  for the axial and polar perturbations around the Reissner--Nordstr\"om solution are related via a duality symmetry, which enforces a degeneracy between the two sectors. At finite frequency this implies  isospectrality properties of the even and odd  modes~\cite{Chandrasekhar:1985kt}, while in the static limit it entails, as we will emphasize below, the equality of the Love numbers.  Such duality, originally derived by Chandrasekhar~\cite{Chandrasekhar:1985kt}, is commonly written in terms of the $Z_{1,2}$ fields  (see section~\ref{sec:axial} and appendix~\ref{app:polarHZ}). Here, we re-express it, in the static limit,  in terms of the gravitational and electromagnetic potentials as~\cite{Pani:2013ija}
\begin{subequations}
\label{eq:dualityC}
\begin{align}
u_{1}(r) & =\frac{\Delta(r)}{r\,\eta(\ell)}\partial_{r}u_{4}(r)-\frac{Q}{r}h_{0}(r) \, , \\ 
H_{0}(r) 
& = \frac{\Delta(r)}{r^2}\partial_r\left[\frac{r^2\,h_0(r)}{\Delta(r)}\right] \, .
\end{align}
\end{subequations}
It is straightforward to check that the polar equations \eqref{eq:polareqs} are linked to the axial ones~\eqref{eq:Cardosocoupled} via the relations \eqref{eq:dualityC}. Note that the transformation \eqref{eq:dualityC}  maps regular solutions at the horizon into regular solutions---recall that the regular branch of $h_0$ scales as $h_0\sim \Delta$ at the Reissner--Nordstr\"om horizon, see eqs.~\eqref{eqh0m1}
 and \eqref{eq:Zihyper}. As a result, it enforces that the linear static response induced by external polar-type perturbations must vanish as a consequence of the vanishing of the odd Love numbers~\cite{Solomon:2023ltn}. 

As an example, we shall use the relations \eqref{eq:dualityC} to write the solutions for $H_0$ and $u_1$ in the case $\ell=2$, starting from  eqs.~\eqref{eq:h0l2} and \eqref{eq:u4l2}:
\begin{align}
\label{eq:H0l2}
H_0^{\ell=2}(r) & = {\cal{A}}_{H_0} r^2
-  2 \left( {\cal{A}}_{H_0} M + {\cal{A}}_{u_1} Q\right) r
+ Q\left( {\cal{A}}_{H_0} Q+4{\cal{A}}_{u_1} M \right) 
- \frac{ 2{\cal{A}}_{u_1} Q^3 }{r} ,
\\
\label{eq:u1l2}
u_1^{\ell=2}(r)
& = {\cal{A}}_{u_1} r^3
- \frac{1}{2}  ( {\cal{A}}_{H_0} Q +6 {\cal{A}}_{u_1} M) (r^2 + Q^2) +\left[ {\cal{A}}_{H_0} M\,Q +2 {\cal{A}}_{u_1} (M^2+Q^2)\right]r
+\frac{{\cal{A}}_{u_1} Q^4}{r} \,  ,
\end{align}
where we redefined the amplitudes of the external polar gravitational and electric potentials as ${\cal{A}}_{H_0}$ and ${\cal{A}}_{u_1}$, respectively. 
Under the assumption that the external source is purely gravitational (i.e., ${\cal{A}}_{u_1}=0$), one finds $H_0^{\ell=2}(r)={\cal{A}}_{H_0}\Delta(r) $ and $ u_1^{\ell=2}(r) =\frac{1}{2}{\cal{A}}_{H_0}Q\Delta(r)$, in agreement with \cite{Cardoso:2017cfl}.

Expressions for generic $\ell$ can be analogously obtained by plugging eqs.~\eqref{eqh0m1} and \eqref{equ4m1} into \eqref{eq:dualityC}: 
\begin{subequations}
\label{u1H0sols}
\begin{align}
    \label{H0sols}
    H_0(r) & = \frac{\Delta(r)}{r^2\eta(\ell)\mu(\ell)^2}  \frac{ \left[r \left(q_1 Z_2(r)-q_2 Z_1(r)\right)\right]''}{(q_1-q_2) } \, , \\ 
    \label{u1sols}
     u_1(r) & = \frac{Q \Delta(r)}{ r\eta (\ell) (q_1-q_2)}\left[Z_1'(r)-Z_2'(r) - \frac{ (r\left(q_1Z_2(r)-q_2Z_1(r)\right)' }{r^2\mu(\ell)^2}\right]  \, ,
\end{align}
\end{subequations}
where $Z_{1,2} $ are given in eq.~\eqref{eq:Zihyper}.

\section{Scalar perturbations on Reissner--Nordstr\"om spacetime}
\label{sec:scalarfield}

In this short section, we briefly recall the equation of a (static) massless spin-0 field around a Reissner--Nordstr\"om black hole. The goal is  to show that, analogously to the gravitational and electromagnetic perturbations, the scalar equation can be also recast in the  form \eqref{eq:HyperG:main}.\footnote{See \cite{Berens:2022ebl} for a thorough study of the ladder symmetries of a scalar field on Reissner--Nordstr\"om.}

Let us start from the Klein--Gordon equation
\begin{equation}
\square \Phi =0 \, ,
\end{equation}
for a scalar field $\Phi$. In the static limit, after introducing the coordinate $z$ in \eqref{eq:defzz} and decomposing $\Phi$ in spherical harmonics as $\Phi(r(z),\theta,\phi) = \sum_{\ell m} S_\ell(z)Y_\ell^m(\theta,\phi)$, the equation for the radial profile  $S_\ell(z)$ takes the hypergeometric form,
\begin{equation}
z(1-z)S_\ell''(z) + (1-2z)S_\ell'(z) + \ell(\ell+1)S_\ell(z) =0 \, .
\label{eqSh}
\end{equation}
It is straightforward to check that the scalar equation \eqref{eqSh} reduces to our master equation \eqref{eq:HyperG:main} after performing the Darboux transformation\footnote{Note that \eqref{eqSh} is a hypergeometric equation in the standard form i.e., $z(1-z)S_\ell''(z) + [c-(a+b+1)z]S_\ell'(z) -abS_\ell(z) =0$,  with parameters $a=-\ell$, $b=\ell+1$ and $c=1$. The Darboux transformation \eqref{eq:DarbouxSF} simply corresponds to a Gauss' contiguous relation, shifting $a\mapsto a-2$ and $b\mapsto b-2$.}
\begin{equation}
    S_\ell(z) = \frac{2 z}{(z-1)^3} \left[ 2 \ell(\ell+1) (z-1)+2 z+1 \right]F'(z)
    +\frac{(\ell+2)(\ell-1)}{(z-1)^3} \left[ \ell(\ell+1) (z-1)+2 z\right] F(z) \, .
\label{eq:DarbouxSF}
\end{equation}

\section{Ladder symmetries of Reissner--Nordstr\"om black holes}
\label{sec:laddersec}

Hypergeometric equations  possess a large set of symmetry properties  leading to well-known and  classified identities. One example of identities is given by contiguous relations connecting hypergeometric functions with parameters differing by integer values.  

In the case of \eqref{eq:HyperG:main}, the contiguous relations of the hypergeometric solutions translate into ladder symmetries connecting physical perturbations of black holes with different multipole number. Such ladder symmetries were introduced in~\cite{Hui:2021vcv}  for spin-$0$, $1$ and $2$ perturbations around Schwarzschild and Kerr black holes, and were later generalized to the case of spin-$0$ field on Reissner--Nordstr\"om spacetime in~\cite{Berens:2022ebl}. One main consequence of the symmetries is that they constrain the static response of black holes and are responsible for the vanishing of the black hole Love numbers in general relativity~\cite{Hui:2021vcv,Hui:2022vbh,BenAchour:2022uqo}. 
In this section, we show that ladder symmetries exist also for electromagnetic and gravitational perturbations of Reissner--Nordstr\"om black holes, completing the program initiated in~\cite{Hui:2021vcv} for non-rotating black holes.

To this end, it is convenient to introduce 
\begin{equation}
    \xi_\ell (z)=(z-1)^{-2}F(z),
\end{equation}
where we made  the $\ell$ dependence explicit in $\xi$. In the following, we will always implicitly assume that $\ell\geq2$.\footnote{Recall that for $\ell=0$ there is no propagating degree of freedom, while for $\ell=1$ only the electromagnetic sector contains a propagating mode. In the latter case, there is only one dynamical equation, and the analysis reduces, in the static limit, to the one of~\cite{Hui:2021vcv}. }
From \eqref{eq:HyperG:main}, $\xi_\ell (z)$ satisfies the equation
\begin{equation}
(1-z)^{-2}\partial_z\left[z (1-z)^5 \partial_z \left(\frac{\xi_\ell (z)}{(z-1)^2} \right)\right]
+ (\ell-2) (\ell+3)  \xi_\ell (z) =0 \, .
\label{eqxi}
\end{equation}
By setting $\ell=2$ in eq.~\eqref{eqxi}, it is immediate to recognize the following conserved quantity for $\xi_2 (z)$:
\begin{equation}
P_2 \equiv z (z-1)^5 \partial_z \left(\frac{\xi_2 (z)}{(z-1)^2} \right) = \text{constant} \, ,
\label{P2}
\end{equation}
which acts as a `horizontal symmetry' for the $\ell=2$ solution and is constant in $z$~\cite{Hui:2021vcv}. The conserved charge $P_2$ is useful to connect asymptotics. To see why, let us start by considering the large-$z$ limit. The two possible falloffs satisfying \eqref{P2} are, at large $z$, $\xi_2^{(1)}\sim z^2$ and  $\xi_2^{(2)}\sim z^{-3}$: the first one has vanishing $P_2$, while the second one corresponds to a solution with nonzero $P_2$. From the conservation of the horizontal charge $P_2$, it is immediate to see that near the horizon (i.e., $z\rightarrow 1$) $\xi_2^{(1)}$ is finite, going as $(z-1)^2$, while $\xi_2^{(2)}$ blows up as $(z-1)^{-2}$. One can then easily show that the divergence of $\xi_2^{(2)}$ is associated to a singularity at the horizon~\cite{Cardoso:2017cfl,Pereniguez:2021xcj}, and therefore that $\xi_2^{(2)}$ is unphysical. 

The existence of the horizontal symmetry $P_2$ is perhaps not surprising, as it is in fact just related to the Wronskian of the differential equation~\cite{Hui:2021vcv,Lisovyy_2022}.\footnote{Let us consider the Wronskian ${\cal W}[\xi_2,\tilde{\xi}_2]\equiv  z(1-z)( \tilde{\xi}_2 \partial_z \xi_2 - \xi_2 \partial_z \tilde{\xi}_2 )$. By construction, ${\cal W}$ is constant, i.e.~$\partial_z {\cal W}=0$, when $\xi_2$ and $\tilde{\xi}_2$ solve \eqref{eqxi}. It is straightforward to check that, up to an irrelevant proportionality constant, $P_2$ in \eqref{P2} coincides with the Wronskian ${\cal W}[\xi_2,\tilde{\xi}_2]$ if $\tilde{\xi}_2$ is taken to be the solution that is regular at the horizon, i.e.~$\tilde{\xi}_2=(z-1)^2$.} More interesting is instead the existence of another set of ladder symmetries, which connect solutions at different level $\ell$ and allow to generalize the previous argument to higher multipoles. Let us rewrite the equation of motion \eqref{eqxi} as
\begin{equation}
{\mathscr{H}}_\ell \xi_\ell =0 \, ,
\qquad 
    {\mathscr{H}}_\ell \equiv z(1-z)\left [(1-z) z \frac{\partial ^2}{\partial z^2}+(1-2 z) \frac{\partial }{\partial z}+ \ell(\ell+1)-\frac{4}{1-z}\right] \, ,
\end{equation}
where ${\mathscr{H}}_\ell$ plays the role of a `Hamiltonian'.
It is straightforward to show that the operators
\begin{subequations}
\label{eq:defDpm}
\begin{align}
     D_\ell^+ & \equiv -z(1-z) \frac{\partial }{\partial z}  + f_\ell(z) \, , \\  
    D_\ell^- & \equiv  z(1-z) \frac{\partial }{\partial z} +  f_{\ell-1}(z) \, ,
\end{align}
\end{subequations}
where 
\begin{equation}
    f_\ell(z)   \equiv (\ell+1) z-\frac{(\ell-1) (\ell+3)}{2 (\ell+1)} \, ,
\end{equation}
satisfy the following algebra with the Hamiltonian ${\mathscr{H}}_\ell$:
\begin{equation} 
    {\mathscr{H}}_{\ell+1}D_\ell^+ = D_\ell^+{\mathscr{H}}_\ell \, , \qquad
    {\mathscr{H}}_{\ell-1}D_\ell^- = D_\ell^-{\mathscr{H}}_\ell \, .
\label{laddercrs}
\end{equation}
From \eqref{laddercrs} it is clear that $D_\ell^\pm$ act as ladder operators: they can be used to generate solutions at levels $\ell\pm1$ from a given solution $\xi_\ell$ at level $\ell$, i.e.~$\xi_{\ell\pm1} = D_\ell^\pm \xi_\ell$. From the definitions \eqref{eq:defDpm}, one can also easily check that the ladder operators satisfy the following relations:
\begin{align}
    D_{\ell+1}^-D_\ell^+ +{\mathscr{H}}_\ell & = \frac{(\ell-1)^2 (\ell+3)^2}{4 (\ell+1)^2}  \, , \\  
    D_{\ell-1}^+D_\ell^-+ {\mathscr{H}}_\ell & = \frac{\left(\ell^2-4\right)^2 }{4 \ell^2}  \, .
\end{align}

The generators $D_\ell^\pm$ are useful because, starting from a seed solution $\xi_{\ell=2}$, we can construct the full tower of solutions for    $\xi_{\ell}$. In particular, they allow to extend the horizontal symmetries to higher $\ell$ and find the connection coefficients for the differential equation \eqref{eqxi}. Note that the coefficients in $D_\ell^\pm$ are smooth functions of $z$ at $z=0,1$: this ensures that $D_\ell^\pm$ map regular solutions at the singular points $z=0,1$ into regular solutions. In particular, this implies that $\xi_\ell^{(1)}$, constructed by acting $\ell-2$ times on $\xi_2^{(1)}$ with $D_\ell^+$, is regular at the horizon $z=1$ and is a finite polynomial in $z$ growing as $\sim z^\ell$ at $z\rightarrow\infty$. Conversely, the solution $\xi_\ell^{(2)}$, generated from the seed $\xi_2^{(2)}$, with decaying falloff at large $z$, is singular at $z=1$.

In conclusion, in analogy with the Schwarzschild case~\cite{Hui:2021vcv}, the vanishing of the Love numbers is a consequence of the fact that the $\xi_\ell$ solution that is regular at the horizon is a polynomial with only positive powers in $z $, while the branch with decaying $1/z^{\ell+1}$ profile is associated with an unphysical divergence  at the black hole horizon.

\section{Matching to point-particle effective field theory}
\label{sec:matching}

Any object, when seen from distances much larger than the its typical size, appears in first approximation as a point source. This intuition is systematized in the point-particle effective theory~\cite{Goldberger:2004jt,Goldberger:2005cd,Goldberger:2009qd,Rothstein:2014sra,Porto:2016pyg,Levi:2018nxp,Goldberger:2020fot,Goldberger:2022ebt,Goldberger:2022rqf}, which takes advantage of the separation of scales in the problem and  provides an organizing principle that  allows to describe, in a robust way, the finite-size  properties of the object, as seen by a far-distance observer. In particular, the EFT provides a conceptually clean definition of tidal Love numbers that is free of ambiguities due to nonlinearities and gauge invariance in the theory, which make the partition into tidal source and response in general relativity less neat, as opposed to Newtonian gravity. 
In addition, the EFT allows to more transparently relate phenomena and observables that are controlled by the same operator coefficient, which might not be so clear in different frameworks. 

In the following, we consider the point-particle EFT of a massive, charged, self-gravitating object. We will neglect dissipative effects and focus on the leading order in the time-derivative expansion, which is  enough to study the static response in the absence of angular momentum. By performing the matching with the full Reissner--Nordstr\"om solution in general relativity of sections~\ref{sec:axial} and \ref{sec:polardual}, we will then derive the Love number couplings. 

The EFT can be written as
\begin{equation}
S_{\rm EFT} = S_{\rm bulk} + S_{\rm pp} + S_{\rm finite-size} \, ,
\label{SEFT0}
\end{equation}
where 
\begin{equation}
S_{\rm bulk} = \frac{1}{16\pi} \int \D^4 x \sqrt{-g}  \left( R - F_{\mu\nu}F^{\mu\nu} \right) \, ,
\end{equation}
is  the Einstein--Maxwell action that describes the dynamics of the gravitational and electromagnetic  fields in the bulk, 
\begin{equation}
S_{\rm pp} =  \int \D \tau \left( - M \sqrt{-g_{\mu\nu} \frac{\D x^\mu}{\D \tau }\frac{\D x^\nu}{\D \tau }}  +Q A_\mu \frac{\D x^\mu}{\D \tau }  \right)
\end{equation}
describes the free motion of the point particle, whose worldline is parametrized by the coordinate  $\tau$, and\footnote{The notation is chosen to mirror the one in~\cite{Hui:2020xxx}.}
\begin{equation}
\begin{split}
S_{\rm finite-size}  = 
    \sum_{\ell=1}^{\infty}  \int \D \tau &\Bigg[ \frac{\lambda_{\ell}^{(E)}}{2\ell!}\left(\partial_{(a_{1}}\cdots\partial_{a_{\ell-1}}E_{a_{\ell})_{T}}\right)^{2}
    +\frac{\lambda_{\ell}^{(B)}}{4\ell!}\left(\partial_{(a_{1}}\cdots\partial_{a_{\ell-1}}B_{a_{\ell})_{T}b}\right)^{2}\\ 
 & + \frac{\lambda_{\ell}^{(C_{E})}}{2\ell!} \left(\partial_{(a_{1}}\cdots\partial_{a_{\ell-2}}E_{a_{\ell-1}a_{\ell})_{T}}^{(2)}\right)^{2}
 +\frac{\lambda_{\ell}^{(C_{B})}}{4\ell!}\left(\partial_{(a_{1}}\cdots\partial_{a_{\ell-2}}B_{a_{\ell-1}a_{\ell})_{T}|b}^{(2)}\right)^{2}
 \\ 
 & +\frac{\eta_{\ell}^{(E)}}{\ell!}\left(\partial_{(a_{1}}\cdots\partial_{a_{\ell-1}}E_{a_{\ell})_{T}}\right)\left(\partial_{(a_{1}}\cdots\partial_{a_{\ell-2}}E_{a_{\ell-1}a_{\ell})_{T}}^{(2)}\right)\\ 
 & +\frac{\eta_{\ell}^{(B)}}{2\ell!}\left(\partial_{(a_{1}}\cdots\partial_{a_{\ell-1}}B_{a_{\ell})_{T}b}\right)\left(\partial_{(a_{1}}\cdots\partial_{a_{\ell-2}}B_{a_{\ell-1}a_{\ell})_{T}|b}^{(2)}\right)\Bigg] 
\end{split}
\label{eq:Sfinitesize}
\end{equation}
captures the leading finite-size effects. The quantities $E$ and $B$ denote the electric and magnetic fields, and the electric and magnetic components of the Weyl tensor $C_{\mu\nu\rho\sigma}$, defined by
\begin{equation}
E_a \equiv F_{0a} = \dot{A}_a - \partial_a A_0 \, ,
\qquad
B_{ab} \equiv F_{ab} = \partial_a A_b - \partial_b A_a \, ,
\end{equation}
\begin{equation}
E^{(2)}_{ab} \equiv C_{0a0b} = - \frac{1}{2}\partial_a\partial_b h_{00} \, ,
\qquad
B^{(2)}_{ab\vert c} \equiv C_{0abc} =  
\frac{1}{2}\left( \partial_a \partial_{b} h_{c0} -\partial_a \partial_{c} h_{b0} \right)
=\partial_a \partial_{[b} h_{c]0} \, ,
\end{equation}
where we linearized in $h_{\mu\nu}$ around the flat spacetime metric, $h_{\mu\nu} \equiv g_{\mu\nu}- \eta_{\mu\nu}$. 
Note that, in the standard spirit of EFTs, in \eqref{eq:Sfinitesize} we added all possible quadratic operators in the fields that are compatible with the symmetries in the theory. In particular, in addition to $\lambda_{\ell}^{(E,B)}$ and $\lambda_{\ell}^{(C_{E,B})}$, which correspond to the Love numbers discussed e.g.~in \cite{Cardoso:2017cfl,Pereniguez:2021xcj}, we have in \eqref{eq:Sfinitesize} also $\eta_{\ell}^{(E,B)}$. Such effective couplings capture the response, due to the object's finite size, induced by the gravitational-electromagnetic mixing. In analogy with $\lambda_{\ell}^{(E,B)}$ and $\lambda_{\ell}^{(C_{E,B})}$, $\eta_{\ell}^{(E,B)}$ can be determined by performing the matching between the EFT and the Reissner--Nordstr\"om solution for the perturbations.

First of all, let us compute, in the effective theory \eqref{SEFT0},  the static response induced by an external, time-independent  gravito-electromagnetic field acting on the point particle. One way of solving this problem is to find the solution of the equation for the response field in the presence of the external source. In practice, we shall decompose the metric and the vector fields as
\begin{equation}
g_{\mu\nu}=\eta_{\mu\nu}+h_{\mu\nu}^{(0)}+h_{\mu\nu}^{(1)},\qquad A_\mu = A_\mu^{(0)} +  A_\mu^{(1)} \, ,
\end{equation}
where $h_{\mu\nu}^{(0)}$ and  $A_\mu^{(0)}$ denote the external gravitational and electromagnetic sources, while $h_{\mu\nu}^{(1)}$ and  $A_\mu^{(1)}$ are the response fields that we want to compute. By construction, $h_{\mu\nu}^{(0)}$ and  $A_\mu^{(0)}$  solve  the bulk (static) equations of motion. Choosing the de Donder gauge for the gravitational sector and the Lorentz gauge for the electromagnetic one i.e., respectively
\begin{equation}
\partial^\mu h_{\mu\nu} - \frac{1}{2} \partial_\nu h =0 \, ,
\qquad
\partial^\mu A_\mu =0 \, ,
\end{equation}
with $h \equiv h^\alpha_\alpha$, a certain number of simplifications occur in the static limit in the bulk, e.g., $\vec \nabla^2 h_a^a = \vec\nabla^2 h_{00} = \vec\nabla^2 h_{0a} = \partial_a h^{a0} = 0$ and $\partial_a h^{ab} = \tfrac{1}{2} \partial^b h_c^c-\tfrac{1}{2}\partial^b h_{00}$, where $\vec\nabla^2\equiv \partial_a\partial^a$ is the spatial laplacian. In the end, the static equations for $h_{\mu\nu}^{(0)}$ and  $A_\mu^{(0)}$ in the  bulk boil down  to 
\begin{equation}
\vec \nabla^2 h_{\mu\nu}^{(0)}(\vec x)=0 \, ,
\qquad
\vec \nabla^2 A_\mu^{(0)}(\vec x)=0\, .
\end{equation}
In the following, it will be enough to focus on the solutions for the components~\cite{Hui:2020xxx}
\begin{equation}
h^{(0)}_{00}(\vec x) = c^{(E)}_{a_1\cdots a_{\ell}}x^{a_1}\cdots x^{a_\ell}\, ,
\qquad
h_{0a}^{(0)}(\vec x) = c^{(B)}_{a | b_1\cdots b_\ell}x^{b_1}\cdots x^{b_\ell}\, ,
\label{tidalh}
\end{equation}
and 
\begin{equation}
A_0^{(0)}(\vec x) = d^{(E)}_{a_1\cdots a_\ell}x^{a_1}\cdots x^{a_\ell} \, ,
\qquad
A_a^{(0)}(\vec x) = d^{(B)}_{a | b_1\cdots b_\ell}x^{b_1}\cdots x^{b_\ell} \, ,
\label{tidalA}
\end{equation}
where $c^{(E)}_{a_1\cdots a_{\ell}}$, $c^{(B)}_{a | a_1\cdots a_\ell}$, $d^{(E)}_{a_1\cdots a_{\ell}}$ and $d^{(B)}_{a | a_1\cdots a_\ell}$ are traceless tensors in the spatial indices $a_1\cdots a_{\ell}$ and correspond to the amplitudes of the external probing fields. 
Using that
\begin{equation}
\partial_{(a_1}\cdots \partial_{a_{\ell-1}}  B_{ a_\ell ) b} =  
 \partial_{a_1}  \cdots  \partial_{a_\ell} A_{b} 
- \frac{1}{\ell}  \left( \partial_{b}\partial_{a_2}  \cdots  \partial_{a_\ell} A_{a_1} + (\ell-1) \times\text{perms.}  \right) \, ,
\end{equation}
\begin{equation}
\partial_{(a_1}\cdots \partial_{a_{\ell-2}}  C_{0 a_{\ell-1} a_\ell ) b} =  
\frac{1}{2} \partial_{a_1}  \cdots  \partial_{a_\ell} h_{b0} 
- \frac{1}{2\ell}  \left( \partial_{b}\partial_{a_2}  \cdots  \partial_{a_\ell} h_{a_10} + (\ell-1) \times\text{perms.}  \right) \, ,
\end{equation}
which imply
\begin{equation}
\left(\partial_{(a_1}\cdots \partial_{a_{\ell-1}}  B_{  a_\ell ) b} \right)^2 =  
2\frac{\ell+1}{\ell }  (  \partial_{a_1}  \cdots  \partial_{[a_\ell} A_{b]} )^2 \, ,
\end{equation}
\begin{equation}
\left(\partial_{(a_1}\cdots \partial_{a_{\ell-2}}  C_{0 a_{\ell-1} a_\ell ) b} \right)^2 =  
\frac{\ell+1}{2\ell }  (  \partial_{a_1}  \cdots  \partial_{[a_\ell} h_{b]0} )^2 \, ,
\end{equation}
\begin{equation}
\left(\partial_{(a_1}\cdots \partial_{a_{\ell-1}}  B_{  a_\ell ) b} \right)\left(\partial_{(a_1}\cdots \partial_{a_{\ell-2}}  C_{0 a_{\ell-1} a_\ell ) b} \right) =  
\frac{\ell+1}{\ell }  (  \partial_{a_1}  \cdots  \partial_{[a_\ell} A_{b]} ) (  \partial_{a_1}  \cdots  \partial_{[a_\ell} h_{b]0} ) \, ,
\end{equation}
and given the solutions \eqref{tidalh} and \eqref{tidalA}, we can extract the response fields $h_{\mu\nu}^{(1)}$ and  $A_\mu^{(1)}$ by solving the inhomogeneous equations 
\begin{subequations}
\label{eq:h1A1}
\begin{align}
\label{eq:h1A11}
\vec \nabla^2 A_0^{(1)}(\vec x)
& = - 2\pi  (-1)^\ell \left( 2 d^{(E)}_{a_1\cdots a_\ell} \lambda_\ell^{(E)} + c^{(E)}_{a_1\cdots a_{\ell}} \eta^{(E)}_\ell  \right) \partial_{a_1} \cdots \partial_{a_\ell} \delta^{(3)}(\vec{x}) \, ,
\\
\label{eq:h1A12}
\vec \nabla^2 A_b^{(1)}(\vec x)
& = - 2\pi  (-1)^\ell \frac{\ell+1}{\ell} \left( 2d^{(B)}_{[b\vert a_1]\cdots a_\ell} \lambda_\ell^{(B)} +    c^{(B)}_{[b\vert a_1]\cdots a_{\ell}} \eta^{(B)}_\ell  \right) \partial_{a_1} \cdots \partial_{a_\ell} \delta^{(3)}(\vec{x}) \, ,
\\
\label{eq:h1A13}
\vec \nabla^2 h_{00}^{(1)}(\vec x)
& = -4\pi  (-1)^\ell \left( c^{(E)}_{a_1\cdots a_\ell} \lambda_\ell^{(C_E)} +2  d^{(E)}_{a_1\cdots a_{\ell}} \eta^{(E)}_\ell  \right) \partial_{a_1} \cdots \partial_{a_\ell} \delta^{(3)}(\vec{x}) \, ,
\\
\label{eq:h1A14}
\vec \nabla^2 h_{0b}^{(1)}(\vec x)
& = -8\pi  (-1)^\ell \frac{\ell+1}{\ell} \left(  c^{(B)}_{[b\vert a_1]\cdots a_\ell} \lambda_\ell^{(C_B)} +   2 d^{(B)}_{[b\vert a_1]\cdots a_{\ell}} \eta^{(B)}_\ell  \right) \partial_{a_1} \cdots \partial_{a_\ell} \delta^{(3)}(\vec{x}) \, .
\end{align}
\end{subequations}

\paragraph{Gravito-magnetic response.} 
Let us start by solving eqs.~\eqref{eq:h1A12} and \eqref{eq:h1A14} for the odd components of the fields.
In Fourier space, we shall write~\cite{Hui:2020xxx}
\begin{subequations}
\label{eq:Abh0bp}
\begin{align}
A_b^{(1)}(\vec p)
 & =  2\pi  (-i)^\ell \frac{\ell+1}{\ell} \left( 2d^{(B)}_{[b\vert a_1]\cdots a_\ell} \lambda_\ell^{(B)} +    c^{(B)}_{[b\vert a_1]\cdots a_{\ell}} \eta^{(B)}_\ell  \right) \frac{p_{a_1} \cdots p_{a_\ell}}{\vec p^2}   \, ,
\\
 h_{0b}^{(1)}(\vec p)
& = 8\pi  (-i)^\ell \frac{\ell+1}{\ell} \left(  c^{(B)}_{[b\vert a_1]\cdots a_\ell} \lambda_\ell^{(C_B)} +  2  d^{(B)}_{[b\vert a_1]\cdots a_{\ell}} \eta^{(B)}_\ell  \right) \frac{p_{a_1} \cdots p_{a_\ell}}{\vec p^2}  \, .
\end{align}
\end{subequations}
Transforming back to real space using the formula
\begin{equation}
(-i)^\ell \int\frac{\D^3 \vec{p}}{(2\pi)^3}\,\e^{i \vec{p}\cdot\vec{x}}\,c^{a_1\cdots a_\ell}\frac{p_{a_1}\cdots p_{a_\ell}}{\vec p^2} = (-1)^\ell\frac{2^{\ell-2}}{\sqrt{\pi}\,\Gamma(\frac{1}{2}-\ell)}\,c_{a_1\cdots a_\ell} \frac{x^{a_1}\cdots x^{a_\ell}}{\vert \vec x \vert^{2\ell+1}} \, ,
\end{equation}
we find
\begin{subequations}
\label{eq:Abh0bx}
\begin{align}
A_b^{(1)}(\vec x)
 & =    (-1)^\ell \frac{\ell+1}{\ell} \frac{2^{\ell-1}\sqrt{\pi}}{\Gamma(\frac{1}{2}-\ell)} \left( 2d^{(B)}_{[b\vert a_1]\cdots a_\ell} \lambda_\ell^{(B)} +    c^{(B)}_{[b\vert a_1]\cdots a_{\ell}} \eta^{(B)}_\ell  \right) \frac{x^{a_1}\cdots x^{a_\ell}}{\vert \vec x \vert^{2\ell+1}}  \, ,
\\
 h_{0b}^{(1)}(\vec x)
& =  (-1)^\ell \frac{\ell+1}{\ell} \frac{2^{\ell+1}\sqrt{\pi}}{\Gamma(\frac{1}{2}-\ell)} \left(  c^{(B)}_{[b\vert a_1]\cdots a_\ell} \lambda_\ell^{(C_B)} +  2  d^{(B)}_{[b\vert a_1]\cdots a_{\ell}} \eta^{(B)}_\ell  \right) \frac{x^{a_1}\cdots x^{a_\ell}}{\vert \vec x \vert^{2\ell+1}}  \, .
\end{align}
\end{subequations}
In order to match these expressions to the full solutions for the gravitational  and  magnetic potentials computed in general relativity,  we have to account for the fact that the two computations have been carried out in different gauges. One option is to explicitly change gauge in one of the answers and then compare. Perhaps more easily, we will instead match the magnetic field and the linearized Weyl tensor, which are  gauge-invariant quantities.
Summing up the tidal field with the induced profiles, and plugging into the definitions of the electromagnetic and Weyl tensors,  we obtain in spherical coordinates, for the angular components,
\begin{subequations}
\label{eq:BFodd}
\begin{align}
B_{ij}
 & \propto  r^{\ell+1}\nabla_{[i}Y_{j] }^{(T)}{}^\ell_m \left[ \bar{d}^{(B)} \left(  1 +(-1)^\ell \frac{\ell+1}{\ell} \frac{2^{\ell}\sqrt{\pi}}{\Gamma(\frac{1}{2}-\ell)} \lambda_\ell^{(B)} r^{-2\ell-1} \right)  
 +\bar{c}^{(B)}(-1)^\ell \frac{\ell+1}{\ell}\frac{2^{\ell-1}\sqrt{\pi}}{\Gamma(\frac{1}{2}-\ell)}    \eta^{(B)}_\ell r^{-2\ell-1} \right] ,
\\
 C_{0rij}
& \propto r^{\ell}\nabla_{[i}Y_{j] }^{(T)}{}^\ell_m  \Bigg[ \bar{c}^{(B)} \left( 1 -  (-1)^\ell  \frac{(\ell+1)(\ell+2)}{\ell(\ell-1)} \frac{2^{\ell+1}\sqrt{\pi}}{\Gamma(\frac{1}{2}-\ell)}  \lambda_\ell^{(C_B)} r^{-2\ell-1}   \right)
\nonumber \\
&\qquad\qquad\qquad\qquad\qquad\qquad\qquad\qquad\qquad\qquad
 -  \bar{d}^{(B)}(-1)^\ell  \frac{(\ell+1)(\ell+2)}{\ell(\ell-1)} \frac{2^{\ell+2}\sqrt{\pi}}{\Gamma(\frac{1}{2}-\ell)}  \eta^{(B)}_\ell  r^{-2\ell-1}   \Bigg] \, ,
\end{align}
\end{subequations}
where we redefined the external amplitudes as $\bar{c}^{(B)}$ and $\bar{d}^{(B)}$, and where $Y_{j }^{(T)}{}^\ell_m$ are the  transverse vector spherical harmonics on the $2$-sphere, which can be written in terms of the Levi--Civita symbol as $Y_{i}^{(T)}{}_\ell^m = \epsilon_{ij}\nabla^j Y_{\ell}^m/\sqrt{\ell(\ell+1)}$~\cite{Hui:2020xxx}.

In order to find the response coefficients $\lambda^{(C_B)}_\ell$, $\lambda^{(B)}_\ell$ and $\eta^{(B)}_\ell$, we shall compare the EFT expressions  \eqref{eq:BFodd} with the full general relativity results. In the notation of appendix~\ref{app:perturbations}, we shall write
\begin{subequations}
\label{eq:BFoddUV}
\begin{align}
B_{ij} & = - \frac{u_4(r)}{\ell(\ell+1)} \nabla_{[i}Y_{j] }^{(T)}{}^\ell_m \, ,
\\
 C_{0rij} & = - \left(h_0'(r)-\frac{2h_0(r)}{r}\right)\nabla_{[i}Y_{j] }^{(T)}{}^\ell_m \, ,
\end{align}
\end{subequations}
where $u_4(r)$ and $h_0(r)$ are the  solutions in the full theory, which can be read off from  our results in  section~\ref{sec:axial}.
Given the explicit solution \eqref{eq:Zihyper} and the considerations in section~\ref{sec:analytic}, we can conclude that $\lambda^{(C_B)}_\ell$, $\lambda_\ell^{(B)}$ and $\eta_\ell^{(B)}$ all vanish for all $\ell$. Note that, for $\ell\geq3$, this  follows straightforwardly from simple dimensional analysis  at the level of the full solutions \eqref{eq:BFoddUV}. The matching of the $\ell=2$ mode requires instead some caution. Indeed, by naively comparing \eqref{eq:BFodd} with \eqref{eq:BFoddUV}, where $h_0$ and $u_4$ are given in \eqref{eq:h0l2} and \eqref{eq:u4l2} respectively, one could have reached a wrong conclusion on the value of $\eta_2^{(B)}$. On the other hand, from our analysis in section~\ref{sec:analytic}, we already know that the decaying falloff in \eqref{eq:h0l2} does not belong to the response series and should therefore not be associated with the Love number coupling $\eta_2^{(B)}$ in \eqref{eq:Sfinitesize}. In section~\ref{sec:analytic}, we reached this conclusion by performing an analytic continuation in $\ell$, and it would be interesting to recover the same result purely at the level of the worldline EFT. Subleading terms in the full general relativity solutions can be reconstructed, order by order in $M$ and $Q$, by computing the one-point functions of $h_0$ and $u_4$  from Feynman diagrams with one external source field and increasing number of insertions on the worldline.  Since the  questioned $1/r^2$ falloff in the full solution  \eqref{eq:h0l2} scales as $Q^5$, we expect it to correspond to a four-loop diagram at order $Q^5$ in the EFT, with five $A_\mu$ insertions on the worldline. We leave this explicit check and the reconstruction of the full tidal field profiles \eqref{eq:h0l2} and \eqref{eq:u4l2} for future work.

\paragraph{Gravito-electric response.} Similarly, we can solve  eqs.~\eqref{eq:h1A11} and \eqref{eq:h1A13} for the even components of the fields. The solutions for the potentials $A_0$ and $h_{00}$ read, respectively,
\begin{subequations}
\label{eq:A0h00x}
\begin{align}
A_0^{(1)}(\vec x)
 & =    (-1)^\ell \frac{2^{\ell-1}\sqrt{\pi}}{\Gamma(\frac{1}{2}-\ell)} \left( 2d^{(E)}_{ a_1\cdots a_\ell} \lambda_\ell^{(E)} +    c^{(E)}_{ a_1\cdots a_{\ell}} \eta^{(E)}_\ell  \right) \frac{x^{a_1}\cdots x^{a_\ell}}{\vert \vec x \vert^{2\ell+1}}  \, ,
\\
 h_{00}^{(1)}(\vec x)
& =   (-1)^\ell \frac{2^{\ell}\sqrt{\pi}}{\Gamma(\frac{1}{2}-\ell)} \left(  c^{(E)}_{ a_1\cdots a_\ell} \lambda_\ell^{(C_E)} +   2 d^{(E)}_{ a_1\cdots a_{\ell}} \eta^{(E)}_\ell  \right) \frac{x^{a_1}\cdots x^{a_\ell}}{\vert \vec x \vert^{2\ell+1}}  \, .
\end{align}
\end{subequations}
 Summing up the tidal field with the solution for the  induced profile, we obtain, up to an overall irrelevant constant, the following expressions for the radial electric field and the linearized Weyl tensor:
\begin{subequations}
\label{eq:EFeven}
\begin{align} 
\label{eq:EFevenE1}
E_{r}
 & \propto  r^{\ell-1}Y^{\ell m} \left[ \bar{d}^{(E)} \left(  1 -(-1)^\ell \frac{2^{\ell}\sqrt{\pi}(\ell+1)}{\ell \, \Gamma(\frac{1}{2}-\ell)} \lambda_\ell^{(E)} r^{-2\ell-1} \right)  
 -\bar{c}^{(E)}(-1)^\ell \frac{2^{\ell-1}\sqrt{\pi}(\ell+1)}{\ell \, \Gamma(\frac{1}{2}-\ell)}    \eta^{(E)}_\ell r^{-2\ell-1} \right] \, ,
\\
 C_{0r0r}
& \propto r^{\ell-2}Y^{\ell m}  \Bigg[  \bar{c}^{(E)} \left( 1 +  (-1)^\ell  \frac{(\ell+1)(\ell+2)}{\ell(\ell-1)}\frac{2^{\ell}\sqrt{\pi}}{\Gamma(\frac{1}{2}-\ell)}  \lambda_\ell^{(C_E)} r^{-2\ell-1}   \right)
\nonumber \\
&\qquad\qquad\qquad\qquad\qquad\qquad\qquad\qquad\qquad
  +  \bar{d}^{(E)}(-1)^\ell  \frac{(\ell+1)(\ell+2)}{\ell(\ell-1)}\frac{2^{\ell+1}\sqrt{\pi}}{\Gamma(\frac{1}{2}-\ell)}  \eta^{(E)}_\ell  r^{-2\ell-1}   \Bigg] \, .
\label{C0r0rEFT}
\end{align}
\end{subequations}
These expressions should be compared with the solutions for the electric field and the Weyl tensor in the full calculation in general relativity.
For the radial component of the electric field we simply have
\begin{equation}
E_r=-\partial_r\left(\frac{u_1(r)}{r}\right)Y^{\ell m} \, ,
\end{equation}
where $u_1$ is given in eq.~\eqref{u1sols}. Recall that $Z_i$ has the form (see eq.~\eqref{eq:Zirsoln})
\begin{equation}
Z_i(r) =
c_{i,\ell}\left\{ \lambda_i(\ell)\left[\frac{24Q^{4}}{\eta(\ell)}\frac{1}{r}-\frac{6Q^{2}q_{i}}{\eta(\ell)}+q_{i}r^{2}-r^{3}(\eta(\ell)-2)\right] + \ldots +O(r^{\ell+1}) \right\} \, ,
\end{equation}
where we wrote explicitly  only the lowest powers in $r$, while the dots contain terms from $r^4$ up to $r^{\ell+1}$. After plugging into~\eqref{u1sols}, it is straightforward to check that $u_1$ is a polynomial in $r$ with only positive powers, except for a $1/r$ term. Comparing the resulting $E_r$ with \eqref{eq:EFevenE1}, this implies, by simple power counting, that  $\eta_\ell^{(E)}=\lambda_\ell^{(E)}=0$.

The $C_{0r0r}$ component of the Weyl tensor can  instead be written as
\begin{equation}
C_{0r0r}= -\frac{1}{2}\partial_r^2 \left(\frac{\Delta }{r^2}H_0(r)\right)+\frac{2Q}{r^2} \partial_r\left(\frac{u_1(r)}{r}\right) \, .
\label{eq:Weyl0r0r_GR}
\end{equation}
Using \eqref{u1H0sols}, after comparing with the EFT result \eqref{C0r0rEFT}, one infers that $\lambda_\ell^{(C_E)}=0$. Note that, similarly to $u_1$, the full solution for $H_0(r)$, obtained from \eqref{H0sols}, is also a finite polynomial with only positive powers except for a $1/r$  term (see eq.~\eqref{eq:H0l2} for an explicit example with $\ell=2$). This implies that $C_{0r0r}$ contains subleading decaying powers of $r$, at most of the form $1/r^5$. Despite resembling an induced $\ell=2$ deformation of the black hole (see eq.~\eqref{C0r0rEFT}), similarly to the axial case above, this term does not correspond to a nonzero Love number, but it should be interpreted as belonging to the subleading tail of  the external source. From a UV perspective, this can be seen by doing an analytic continuation in $\ell$ in the equations of $Z_1$ and $Z_2$, as we did in section~\ref{sec:analytic}, and by using \eqref{u1H0sols}. Within the EFT, such falloff can be recovered by computing a two-loop diagram, which  corresponds to the order-$Q^3$ correction to the $H_0$ tidal field solution.

\section{Conclusions}

Black holes are notoriously among the simplest macroscopic objects in nature. As it is well known, asymptotically flat solutions in general relativity are completely fixed in terms of just three parameters: mass, spin and charge. Such uniqueness and simplicity of the solution is in turn inherited by the perturbations, whose dynamics is highly constrained by the symmetries in the theory. Some examples include isospectrality and the vanishing of the Love numbers. The former means that, despite a gravitational wave carries two helicity modes, there is in the end a single degenerate observable spectrum of emission, as a consequence of a duality symmetry that relates them~\cite{Chandrasekhar:1975nkd,Chandrasekhar:1975zza,Chandrasekhar:1985kt,Berti:2009kk,Brito:2013yxa,Rosen:2020crj}. In addition, hidden ladder symmetries in general relativity establish that the tidal response coefficients of an asymptotically flat, four dimensional black hole must vanish identically \cite{Hui:2021vcv,Hui:2022vbh}, making the black hole, to a far-distance observer, indistinguishable from a point particle, as long as static perturbations are concerned. 

Although there are good reasons to expect that astrophysical black holes are neutral (e.g., \cite{Gibbons:1975kk}), studying charged solution turns out to be a very useful laboratory to clarify the special property of general relativity and shed light on the symmetry structure of gravity. 
Our main results here can be summarized as follows. First, we proved in full generality that, in the static limit, perturbations of  Reissner--Nordstr\"om black holes are described by a single, decoupled master hypergeometric equation, which captures both gravitational and electromagnetic, as well as spin-0, perturbations. This has been obtained through a nontrivial (generalized) Darboux-like field redefinition that brings the number of singularities in the original  Fuchsian differential equation down to three.
In the case of the axial sector, our result agrees with the findings of \cite{Pereniguez:2021xcj}, although it holds more in general  and provides a more explicit and systematic recipe for the removal of the singular points. 
The master equation allows in the end to more easily derive general solutions, with respect to previous formulations in the literature, valid for generic $\ell$, and unify at the same time different spins.

Second,  we pointed out a subtlety in the identification of the tidal response coefficients that correspond to mixed quadratic operators in the point-particle EFT, which were not considered  in previous analyses~\cite{Cardoso:2017cfl,Pereniguez:2021xcj}.  Background nonlinearities of the Reissner--Nordstr\"om solution, in addition to inducing gravitational-electromagnetic mixing, are also responsible for generating a decaying tail in the full solution of the tidal source. By performing an analytic continuation, we showed explicitly that such a falloff should not be ascribed to the black hole’s response—as a naive power counting might have erroneously led to think—confirming that the Reissner--Nordstr\"om Love number couplings, including those of mixed type, vanish in four spacetime dimensions. 

Finally, we showed that underlying the Reissner--Nordstr\"om perturbations is a hidden structure of ladder symmetries, which is responsible for the explicit form of the static solutions and the vanishing of the tidal response. 

We envision various directions that will be worth exploring in the future on the subject. 
First, it would be interesting to check which of the results derived here---for instance the removal of the singular points or the ladder symmetries---can be extended to  higher $D$-dimensions, in particular to the gravitational scalar and electric sectors~\cite{Pereniguez:2021xcj,Charalambous:2024tdj}. It is worth recalling that, even though there is no known Chandrasekhar duality for Schwarzschild--Tangherlini black holes in $D>4$, the Heun equation for the gravitational scalar (i.e., even) perturbations can still be recast into hypergeometric form~\cite{Kodama:2003jz,Ishibashi:2003ap,Hui:2020xxx}, and it would be interesting to see whether this remains true in the presence of charge. In addition, it would be interesting to explicitly perform the matching with the point-particle EFT for charged black holes in $D>4$~\cite{Kol:2011vg,Hui:2020xxx,Hadad:2024lsf}, including spin~\cite{Rodriguez:2023xjd,Charalambous:2023jgq} as well as magnetic charge~\cite{Pereniguez:2023wxf}. To date, the static response of Kerr--Newman black holes, for generic spin values, is yet unknown. One difficulty to analytically computing it is that the equations for the Kerr--Newman perturbations do not seem to be separable---not even in the static limit (e.g., \cite{Dias:2022oqm}). In the future, it will interesting to see  whether one can find a formulation that allows to simplify the static equations and obtain an analytic formula for the response coefficients.

\vspace{-.3cm}
\paragraph{Acknowledgements:} We thank Roman Berens and David Pere\~niguez for comments on the draft. 
The work of LS was supported by the Programme National GRAM of CNRS/INSU with INP and IN2P3 co-funded by CNES. The work of MR was supported by DOE Grant DE-SC0010813. MR thanks ICTP, Trieste for hospitality while this work was being initiated.

\appendix
\section{Linearized perturbations of Reissner--Nordstr\"om black holes}
\label{app:perturbations}

In this appendix, we briefly review some details and notation of the linearized equations for the perturbations of a Reissner--Nordstr\"om black hole.
The background Reissner--Nordstr\"om metric $\bar{g}_{\mu\nu}$ and   Maxwell 4-potential $\bar{A}_\mu$ can be read off from eqs.~\eqref{gmunuRN} and \eqref{Amu} respectively. The tensor and vector perturbations, $\delta g_{\mu\nu}= g_{\mu\nu}-\bar{g}_{\mu\nu}$ and $\delta A_\mu = A_\mu- \bar{A}_\mu$, around the background solution can be conveniently separated in parity-even (polar) and parity-odd (axial) parts, and can be parametrized as (see, e.g., \cite{Chandrasekhar:1985kt,Pani:2013ija,Pani:2013wsa,Cardoso:2017cfl,Hui:2020xxx,Pereniguez:2021xcj})
\begin{equation}
\begin{split}
\label{eq:hmunupert_GR}
\delta g_{\mu\nu}= &
\left(\begin{array}{cccc}
\frac{\Delta(r)}{r^2} H_0^{\ell m}(t,r)Y^{\ell m} & {\cal H}_1^{\ell m}(t,r)Y^{\ell m} & 0 & 0 \\
 {\cal H}_1^{\ell m}(t,r)Y^{\ell m} &  \frac{r^2}{\Delta(r)} {\cal H}_2^{\ell m}(t,r)Y^{\ell m} & 0 &0\\
 0 &0 & r^2 K^{\ell m}(t,r)Y^{\ell m}& 0\\ 
0&0 &0 & r^2\sin^2{\theta} K^{\ell m}(t,r)Y^{\ell m}
\end{array}\right)  
\\
 & + \left(\begin{array}{cccc}
0 & 0 & h_0^{\ell m}(t,r)S_\theta^{\ell m} & h_0^{\ell m}(t,r)S_\varphi^{\ell m} \\
0 &0 & h_1^{\ell m}(t,r)S_\theta^{\ell m} & h_1^{\ell m}(t,r)S_\varphi^{\ell m} \\
h_0^{\ell m}(t,r)S_\theta^{\ell m} &h_1^{\ell m}(t,r)S_\theta^{\ell m}  &0 &0\\
h_0^{l\ell m}(t,r)S_\varphi^{\ell m}& h_1^{\ell m}(t,r)S_\varphi^{\ell m} & 0 &0 \end{array}
\right) \,  ,
\end{split}
\end{equation}
and 
\begin{equation}
\delta A_{\mu}=\left(\frac{u_1^{\ell m}(r,t)}{r} Y^{\ell m},\frac{ u_2^{\ell m}(r,t)}{\Delta(r)/r} Y^{\ell m} ,\frac{u_3^{\ell m}(r,t) }{\eta(l)} Y_{b}^{\ell m} \right)
+ \left(0,0,\frac{u_4^{\ell m}(r,t) }{\eta(\ell )}S_{b}^{\ell m}\right)\, ,\label{eq:EMpert_GR}
\end{equation}
where we decomposed in spherical harmonics  $Y^{\ell m}(\theta,\varphi)$, and where we introduced the notations $Y_{b}^{\ell m}\equiv(Y_{,\theta}^{\ell m},Y_{,\varphi}^{\ell m})$ and $S_b\equiv\left(S_\theta^{\ell m},S_\varphi^{\ell m}\right)\equiv (-Y_{,\varphi}^{\ell m}/\sin{\theta},\,\sin{\theta} \,Y_{,\theta}^{\ell m})$.
In \eqref{eq:hmunupert_GR} we fixed the Regge--Wheeler gauge \cite{Regge:1957td}, which allows to remove four independent components in $\delta g_{\mu\nu}$. Gauge invariance in the spin-1 sector allows in addition to set  $u_3(r,t)=0$ in $\delta A_\mu$.

\subsection{Linearized perturbations: odd sector}

Plugging \eqref{eq:hmunupert_GR} and \eqref{eq:EMpert_GR} into the Einstein--Maxwell equations, after some manipulations one can derive the equations for the propagating degrees of freedom. For instance, in the odd sector, one finds, in the static limit, the system of equations~\cite{Cardoso:2017cfl,Pani:2013wsa,Pani:2013ija}:
\begin{subequations}
\label{eq:Cardosocoupled}
\begin{align}
\label{eq:Cardosocoupled1}
\Delta(r)\left[r^{2}h_{0}''(r)-\frac{4Q}{\eta(\ell)}u_{4}'(r)\right]-h_{0}(r)\left[r(r\,\eta(\ell)-4M)+2Q^{2}\right] & =0 \, ,
\\
\label{eq:Cardosocoupled2}
\Delta (r) u_{4}''(r)+2\left(M-\frac{Q^{2}}{r}\right)u_{4}'(r)-\eta(\ell)\,u_{4}(r)+Q\,\eta(\ell)\left[\frac{2h_{0}(r)}{r}-h_{0}'(r)\right] & =0 \, ,
\end{align}
\end{subequations}
which couples the spin-2 and spin-1 degrees of freedom. It is straightforward to check that, with the identification 
\begin{subequations}
\label{eq:CardosotoChandra}
\begin{align}
\label{eq:CardosotoChandra1}
    u_{4}(r)&=Q\,H_{1}^-(r)\, , \\ 
     h_{0}(r)&=\frac{2Q\,\Delta(r)}{\eta(\ell)\,\mu(\ell)\,r^{2}}(rH_{2}^-(r))'  \, ,
\end{align}
\end{subequations}
the coupled equations \eqref{eq:Cardosocoupled} are compatible with \eqref{eq:ChandraH1H2} \cite{Chandrasekhar:1985kt}.\footnote{To stress the difference from the polar fields in the subsection below, in eqs.~\eqref{eq:CardosotoChandra} we restored the $-$ superscript in $H_1^-$ and $H_2^-$, which we had dropped  in eqs.~\eqref{eq:ChandraH1H2} for ease of notation.}

\subsection{Linearized perturbations: even sector}
\label{app:polarHZ}

The polar equations can be derived analogously and take the form
\begin{subequations}
\begin{align}
\frac{\D}{\D r}\left[ \Delta(r) H_{0}'(r) \right] +\left[\frac{4(Q^{2}-M^{2})}{\Delta(r)}-\frac{2Q^{2}}{r^{2}} -\eta(\ell) \right] H_{0}(r)
{-}\frac{4Q}{r}\left[u'_{1}(r)+\frac{\left(Q^{2}-r^{2}\right)u_{1}(r)}{r\,\Delta(r)}\right] &  =0 \, ,\\
 u_{1}''(r)+\frac{\frac{4Q^{2}}{r^{2}}-\eta(\ell)}{\Delta(r)}u_{1}(r){-}\frac{Q}{r}\left[H'_{0}(r)-\frac{2\left(Q^{2}-Mr\right)}{r\Delta(r)}H_{0}(r)\right] & =0 \, .
\end{align}
\label{eq:polareqsAPP}
\end{subequations}
A different, equivalent formulation of the polar equations is given by~\cite{Chandrasekhar:1985kt}:
\begin{subequations}
\label{eq:ChandraH1H2polar}
\begin{align}
 r^2\frac{\D}{\D r}\left[ \frac{\Delta}{r^2}   \frac{\D}{\D r}  H_{2}^+(r)\right]  -\frac{1}{r}\left[ U(r) H_2^+(r)  + W(r) ( -3M H_2^+(r) +2 Q \mu (\ell) H_1^+(r) )\right] & =0 \, , 
\label{eq:ChandraH1H21polar}
 \\ 
 r^2\frac{\D}{\D r}\left[ \frac{\Delta}{r^2}  \frac{\D}{\D r} H_{1}^+(r)\right] -\frac{1}{r}\left[ U(r) H_1^+(r)  + W(r) ( 3M H_1^+(r) +2 Q \mu (\ell) H_2^+(r) )\right] & =0 \, ,
\label{eq:ChandraH1H22polar}
\end{align}
\end{subequations}
where $H_1^+$ and $H_2^+$ denote the polar fields and
\begin{align}
U(r) &\equiv \left(  \mu^2(\ell) r + 3M  \right) W(r) +  \varpi- \frac{\mu^2(\ell)r}{2} -M  +\frac{ \mu^2(\ell)\Delta(r)}{\varpi} \, ,
\\
W(r) & \equiv \frac{\Delta(r)}{r\varpi^2}\left( \mu^2(\ell) r + 3M  \right) +\frac{1}{\varpi}\left( \frac{\mu^2(\ell)r}{2} +M \right)  ,
\end{align}
with $\varpi\equiv \frac{1}{2}\mu^2(\ell)r+3M-\frac{2}{r}Q^2$. We refer to~\cite{Chandrasekhar:1985kt} for details. Using the same field redefinition \eqref{fr12},
\begin{subequations}
\label{fr12polar}
\begin{align}
     H_1^+(r) & =\frac{ Z_1^+(r)- Z_2^+(r)}{q_1- q_2} \, ,
     \label{fr1}
    \\ 
    H_2^+(r) & = \frac{1}{2Q\mu} \frac{q_1 Z_2^+(r) - q_2 Z_1^+(r)}{q_1- q_2}\, ,
    \label{fr2}
\end{align}
\end{subequations}
the  equations for $H_1^+$ and $H_2^+$ decouple, and one finds
\begin{subequations}
\begin{align}
r^2\frac{\D}{\D r}\left[ \frac{\Delta(r)}{r^2}  \frac{\D}{\D r} Z_{1}^+(r)\right] -\frac{1}{r} \left[U(r) + \frac{1}{2} \left( q_1-q_2 \right) W(r) \right] Z_1^+(r) & =0 \, , 
\\
r^2\frac{\D}{\D r}\left[ \frac{\Delta(r)}{r^2}  \frac{\D}{\D r} Z_{2}^+(r)\right] -\frac{1}{r} \left[U(r) - \frac{1}{2} \left( q_1-q_2 \right) W(r) \right] Z_2^+(r) & =0 \, .
\end{align}
\label{eq:Z12polar}
\end{subequations}
In terms of $z$, defined in \eqref{eq:defzz}, the polar equations \eqref{eq:Z12polar} can be equivalently rewritten as
\begin{multline}
    0=\partial_z^2 Z^+_i +\frac{(2z\mathcal{R}+z-\mathcal{R})}{z(z-1)(z+\mathcal{R})} \partial_z Z^+_i + 
     \Bigg[\frac{\mathcal{R}-z (2 \mathcal{R}+1)}{(z-1) z (z+\mathcal{R})^2}-\frac{2 (\eta -2) M}{(2 \mathcal{R}+1) w(z) (z+\mathcal{R})}
     \\ 
      -\frac{M \left(z \left(6 z-4 z^2-3\right)+\eta ^2 (z+\mathcal{R})^3-\mathcal{R}\right) (M (4 (\eta -2) z+4 (\eta +1) \mathcal{R}+6)+(2 \mathcal{R}+1) (q_i-q_j))}{4 (z-1) z (2 \mathcal{R}+1)^2 w(z)^2 (z+\mathcal{R})^2}\Bigg]Z_i^+,
     \label{eq:polarZ}
\end{multline}
where $i,j=1,2$ with $j\neq i$, and where  
\begin{equation}
w(z) = {-}\frac{ M \left(\eta  (z+\mathcal{R})^2+z (-2 z+2 \mathcal{R}+3)-\mathcal{R}\right)}{(2 \mathcal{R}+1) (z+\mathcal{R})} \, .
\label{eq:defwe}
\end{equation}
Note that, with respect to the the axial sector (see eq.~\eqref{eq:normalpsi}), the equations \eqref{eq:polarZ} for the polar fields $Z_i^+$ have two more singularities, corresponding to the roots of $w(z)$ in \eqref{eq:defwe}.
Instead of going through the Fuchsian analysis of appendix~\ref{app:masterequation}, a more convenient way to study the static response of polar-type perturbations is to take advantage of the even-odd duality,  discussed in section~\ref{sec:polardual}.

\section{Reissner--Nordstr\"om perturbations and removable singularities}
\label{app:masterequation}

In this appendix we focus on the equation~\eqref{eq:normalpsi} describing the axial spin-1 and spin-2 perturbations of a Reissner--Nordstr\"om black hole. The equation has four regular singularities in the variable $z$ at the locations $\{-\mathcal{R},0,1,\infty\}$, with $\mathcal{R}>0$. As such it belongs to the Heun class. Via the rescaling 
\begin{equation}
Z_i(r(z))=z^{-\frac{1}{2}}(1-z)^{-\frac{1}{2}}(z+\mathcal{R})\psi_i(z) \, ,
\label{eq:phipsi}
\end{equation}
it can be brought to the Heun normal form
\begin{equation}
\psi_i''(z)+\left[\frac{1-\text{\ensuremath{\alpha_{1}^2}}}{4z^{2}}+\frac{1-\text{\ensuremath{\alpha_{2}^2}}}{4(z-1)^{2}} +\frac{1-\text{\ensuremath{\alpha_{3}^2}}}{4(z+\mathcal{R})^{2}} +\frac{\text{\ensuremath{\beta_{1}}}}{z}+\frac{\text{\ensuremath{\beta_{2}}}}{z-1}+\frac{\text{\ensuremath{\beta_{3}}}}{z+\mathcal{R}}\right]\psi_i(z)  =0 \, ,
\label{heunnormaleq}
\end{equation}
where the parameters are
\begin{equation}
\label{alpha123}
\text{\ensuremath{\alpha_{1}^2}}=0 \, , \qquad \text{\ensuremath{\,\alpha_{2}^2}}=0 \, , \qquad
\text{\ensuremath{\alpha_{3}^2}}=25 \, ,
\end{equation}
and
\begin{subequations}
\label{beta123}
\begin{align}
    \beta_1 & =\eta-\frac{q_j(2 \mathcal{R}+1)}{2 M \mathcal{R}}+\frac{5}{\mathcal{R}}+\frac{9}{2}\, ,
\\
 \beta_2 & = -\eta+\frac{q_j (2 \mathcal{R}+1)}{2 M (\mathcal{R}+1)}+\frac{5}{\mathcal{R}+1}-\frac{9}{2}\, ,
\\
 \beta_3 & = -\frac{(2 \mathcal{R}+1) (10 M-q_j)}{2 M \mathcal{R} (\mathcal{R}+1)}\, ,
\end{align}
\end{subequations}
which satisfy the standard constraint $\beta_1+\beta_2+\beta_3=0$.

Note that $\alpha_3^2$ in \eqref{alpha123} is a positive integer number. As we shall see, this is associated to the fact that the singularity at $z=-\mathcal{R}$ can be removed via a suitable field redefinition. A discussion on how to test whether a given singularity is apparent can be found e.g.~in \cite{Cui_2016,eremenko2018fuchsian}. 
Operationally, we shall first rewrite the equation \eqref{heunnormaleq}  by expanding the coefficient of $\psi_i(z)$ in series around $z=-\mathcal{R}$:
\begin{equation}
(z+\mathcal{R})^{2}\psi_i''(z)+\left(\sum_{k=0}^{\infty}x_{k}(z+\mathcal{R})^{k}\right)\psi_i(z)=0 \, ,
\end{equation}
where it is convenient to rewrite the coefficients $x_k$ of the Taylor series as 
\begin{equation}
x_k \equiv \frac{1-l^2_k}{4} \, ,
\end{equation}
where in particular $l_0\equiv \alpha_3$.
The singular point $z=-\mathcal{R}$  is apparent if and only if $l_0=\alpha_3$ is a positive integer and the following determinant vanishes~\cite{eremenko2018fuchsian}:
\begin{equation}
\label{eq:detapp}
Y_{l_0}(x_{1},x_{2},\cdots,x_{l_0})\equiv \text{det}\left[\begin{array}{ccccc}
x_{1} & 1(1-l_0) & 0 & \cdots & 0\\
x_{2} & x_{1} & 2(2-l_0) & \cdots & 0\\
\vdots & \vdots & \vdots & \ddots & \vdots\\
x_{l_0-1} & x_{l_0-2} & x_{l_0-3} & \cdots & (l_0-1)(-1)\\
x_{l_0} & x_{l_0-1} & x_{l_0-2} & \cdots & x_{1}
\end{array}\right] =0 \, .
\end{equation}
In our case, $\alpha_3=5$ and  
\begin{subequations}
\label{xes}
\begin{align}
x_{0} & =-6  \, ,
\\
x_{1} &=\beta_{3} \, ,
\\
x_{2}& =\frac{1}{4}\left(\frac{1-\alpha_{1}^2}{\mathcal{R}^{2}}+\frac{1-\text{\ensuremath{\alpha_{2}^2}}}{(\mathcal{R}+1)^{2}}-\frac{4\text{\ensuremath{\beta_{1}}}}{\mathcal{R}}-\frac{4\text{\ensuremath{\beta_{2}}}}{\mathcal{R}+1}\right) \, ,
\\ 
 x_{3}&=\frac{1}{2}\left(\frac{1-\alpha_{1}^2}{\mathcal{R}^{3}}+\frac{1-\text{\ensuremath{\alpha_{2}^2}}}{(\mathcal{R}+1)^{3}}-\frac{2\text{\ensuremath{\beta_{1}}}}{\mathcal{R}^2}-\frac{2\text{\ensuremath{\beta_{2}}}}{(\mathcal{R}+1)^2}\right) \, , 
 \\ 
x_{4}&=\frac{1}{4}\left(\frac{3(1-\alpha_{1}^2)}{\mathcal{R}^{4}} +\frac{3(1-\text{\ensuremath{\alpha_{2}^2}})}{(\mathcal{R}+1)^{4}}  -\frac{4\text{\ensuremath{\beta_{1}}}}{\mathcal{R}^{3}}-\frac{4\text{\ensuremath{\beta_{2}}}}{(\mathcal{R}+1)^{3}}\right) \, , 
\\
x_{5}&=\frac{1}{4}\left(\frac{4(1-\text{\ensuremath{\alpha_{1}^2}})}{\mathcal{R}^{5}} +\frac{4(1-\text{\ensuremath{\alpha_{2}^2}})}{(\mathcal{R}+1)^{5}} -\frac{4\text{\ensuremath{\beta_{1}}}}{\mathcal{R}^{4}}-\frac{4\text{\ensuremath{\beta_{2}}}}{(\mathcal{R}+1)^{4}}\right) \, ,
\end{align}
\end{subequations}
and it is straightforward to check that, with the values \eqref{xes}, the condition \eqref{eq:detapp} is indeed satisfied:
\begin{equation}
Y_{5}(x_{1},x_{2},x_{3},x_{4},x_{5})=0 \, .
\end{equation}

Having established that $z=-\mathcal{R}$  is an apparent singularity, one can explicitly find the transformation that removes it and recasts the  equation  into a new one with one less singular point. To this end, it is convenient to start from the Heun standard form:\footnote{The parameter $m$  in \eqref{eqapp:standardheun} should not be confused with the magnetic quantum number of the spherical harmonics $Y^{\ell m}(\theta,\varphi)$ used in the previous sections.}
\begin{equation}
y_i''(z)-\left(\frac{\alpha}{z}+\frac{\beta}{z-1} + \frac{m}{z+\mathcal{R}}\right) y_i'(z) +\frac{d_{0}+\gamma\delta z}{z(z-1)(z+\mathcal{R})} y_i(z) =0 \, ,
\label{eqapp:standardheun}
\end{equation}
with $\alpha +\beta+m+\gamma+ \delta+1=0$, and 
where $y_i(z)$ is related to the normal-form  $\psi_i(z)$  via
\begin{equation}
\psi_i(z)=z^{-\frac{\alpha}{2}}(z-1)^{-\frac{\beta}{2}}(z+\mathcal{R})^{-\frac{m}{2}} y_i(z) \, .
\label{eq:psiy}
\end{equation}
The parameters in \eqref{eqapp:standardheun} are related to the ones in \eqref{heunnormaleq} via the standard mapping:
\begin{equation}
\frac{1}{4} (1-\alpha_1^2)=\frac{\alpha}{2}  \left(-\frac{\alpha }{2}-1\right), 
\quad
\frac{1}{4} (1-\alpha_2^2)=\frac{\beta}{2} \left(-\frac{\beta }{2}-1\right),
\quad
\frac{1}{4} (1-\alpha_3^2)=\frac{m}{2}  \left(-\frac{m}{2}-1\right),
\end{equation}
\begin{equation}
\beta_1 =\frac{\alpha  \beta }{2}-\frac{d_0}{\mathcal{R}}-\frac{\alpha  m}{2 \mathcal{R}},
\quad
\beta_2=-\frac{\alpha  \beta }{2}+\frac{\gamma \delta + d_0}{\mathcal{R}+1} - \frac{\beta  m}{2 (\mathcal{R}+1)},
\quad
\beta_3=-\beta_2-\beta_1 .
\end{equation}
Therefore, in our case, we find
\begin{equation}
\alpha=-1\, ,
\qquad
\beta=-1\, ,
\qquad
m=4\,,
\qquad
\gamma\delta=2-\eta(\ell) \, ,
\qquad
d_0=-\frac{q_i(2  \mathcal{R}+1)}{2 M}-(\eta-2) \mathcal{R} \, .
\end{equation}
Given the standard form \eqref{eqapp:standardheun}, the transformation that removes $z=-\mathcal{R}$ should be looked for in the form $y(z)={\cal Q}(z\frac{\D}{\D z})F(z)$, where ${\cal Q}(z\frac{\D}{\D z})$ is a polynomial of the differential operator $z\frac{\D}{\D z}$, with constant coefficients, of degree fixed by the characteristic exponent at the singularity $z=-\mathcal{R}$~\cite{eremenko2018fuchsian}. In the present case, the degree of the polynomial is $m=4$ and the field redefinition takes the form 
\begin{equation}
y_i(z)=\left[\left(z\frac{\D}{\D z}\right)^{4}+A_i\left(z\frac{\D}{\D z}\right)^{3}+B_i\left(z\frac{\D}{\D z}\right)^{2}+C_i\left(z\frac{\D}{\D z}\right)+D\right]F(z) \, ,
\label{eq.yhyp}
\end{equation}
where $A_i$, $B_i$, $C_i$ and $D$ are (with $i,j=1,2$ and $j\neq i$)
\begin{subequations}
\label{eq:ABCDi}
\begin{align}
     A_i & = -\frac{3 (q_i+2 q_j (\mathcal{R}+1))}{(\mathcal{R}+1) (q_i+ q_j)} \, , \\ 
    B_i & = \frac{q_i (2-7 \mathcal{R})+11 q_j (\mathcal{R}+1)}{(\mathcal{R}+1) (q_i+q_j)} \, , \\ 
    C_i & = -\frac{3 \left(q_i \mathcal{R} (\eta+2 (\eta-1) \mathcal{R}-2)+2 q_j (\mathcal{R}+1)^2\right)}{(\mathcal{R}+1)^2 (q_i+q_j)} \, , \\ 
    D  & = -\frac{ \eta\,\mu^2\, \mathcal{R}^2}{(\mathcal{R}+1)^2} \, .
\end{align}
\end{subequations}
It is straightforward to check that \eqref{eqapp:standardheun}, with \eqref{eq.yhyp} and \eqref{eq:ABCDi}, is 
 equivalent to
\begin{equation}
z(1-z)F''(z)+(1+2z)F'(z)+[\eta(\ell)-2]F(z)=0 \, ,
\label{eq:HyperG}
\end{equation}
which is indeed a hypergeometric equation with singularities at $z=0,1,\infty$. The most general solution to the hypergeometric equation \eqref{eq:HyperG} is given by the linear superposition \eqref{eq:sol2F1}. Then, as we argued in the main text, requiring regularity of the solution at the black hole event horizon fixes the integration constants $d_\ell$ to zero. The physical solution for $Z_i$ can then be cast in the form \eqref{eq:Zihyper}. In the following, we provide an equivalent alternative way of writing the solution \eqref{eq:Zihyper} for $Z_i$.

Using standard derivative identities for the hypergeometric function and the explicit expressions \eqref{eq:ABCDi} for the constants of $A_i$, $B_i$, $C_i$ and $D$, we shall rewrite the physical solution for $Z_i$ as
\begin{align}
 \label{eq:phizsoln}
    Z_i(r(z))  
   & = \frac{c_{i,\ell}\, \eta \,(\eta-2)  }{24 (\mathcal{R}+1)^2 (z+\mathcal{R})} \Biggl[ 12 \,\mathcal{R}(2\mathcal{R}+1)\,\frac{q_i}{M} \, z\hypergeom{2}{1}(-\ell-1,\ell,2;z) -24 \mathcal{R}^2 \hypergeom{2}{1}(-\ell-2,\ell-1,1;z) \nonumber \\ \nonumber 
   &\quad +\eta z^3 (\mathcal{R}+1) \Bigl(\left(\eta -2\right)  (\mathcal{R}+1)\,z \hypergeom{2}{1}(2-\ell,\ell+3,5;z) - 2\,(2\mathcal{R}+1)\,\frac{q_i}{M} \hypergeom{2}{1}(1-\ell,\ell+2,4;z)\Bigr)
   \Biggr] \\ \nonumber 
   & =\frac{c_{i,\ell} \,  \eta\, (\eta-2)  }{24 (\mathcal{R}+1)^2 (z+\mathcal{R})}\sum _{m=0}^{\ell+2} (-z)^m \Biggl[ \Theta(m-4)\eta(\eta -2)(\mathcal{R}+1)^2\, {\ell-2 \choose m-4} \frac{  (\ell+3)_{m-4}}{ (5)_{m-4} } \\ \nonumber 
   &\quad -\frac{\Theta(m-1)\,12 \,\mathcal{R}(2\mathcal{R}+1)\,q_i}{M}\, {\ell+1 \choose m-1}\frac{ (\ell)_{m-1}}{ (2)_{m-1} }-24 \mathcal{R}^2\,{\ell+2 \choose m}\frac{ (\ell-1)_m}{ (1)_m } \\ 
   &\quad  +\frac{\Theta(m-3)\,2 q_i\,\eta  (\mathcal{R}+1)(2\mathcal{R}+1)}{M} \,{\ell-1 \choose m-3}\,\frac{ (\ell+2)_{m-3}}{ (4)_{m-3} } \Biggr]
\end{align}
where $q_{1,2}$ are defined in eqs.~\eqref{eq:q1q2}, ${n \choose x}=\frac{n!}{x!(n-x)!}$ is the standard binomial coefficient,  $(a)_n=\frac{\Gamma(a+n)}{\Gamma(a)}$ is the Pochhammer symbol, and $\Theta(n)$ is the (discrete) Heaviside step function, $\Theta(n)=1$ for all $n\geq0$. 

As argued in section~\ref{sec:masterZ}, the series representation of the physical solution for $Z_i$ is finite. In particular,  $(z+\mathcal{R})Z_i(r(z))$ is a finite polynomial in $z$ of degree $(\ell+2)$. Since $z + \mathcal{R}=\frac{r}{2\sqrt{M^2-Q^2}}=r\,\frac{(1 + 2 \mathcal{R})}{2 M}$,   $rZ_i(r)$ is also a polynomial of degree $(\ell+2)$ in $r$. Explicitly, we shall rewrite it as 
\begin{equation}
\begin{split}
   Z_i(r)  & =\frac{c_{i,\ell}\,\eta\,(\eta-2)\,  M }{12\,r\, (\mathcal{R}+1)^2(1 + 2 \mathcal{R}) }\sum _{m=0}^{\ell+2} \alpha_{m,i}\,\mathcal{R}^{m}\,\sum _{a=0}^{m}\, {m \choose a}\,r^a\left(\frac{1 + 2 \mathcal{R}}{2 M}\right)^a \\
    & = \frac{c_{i,\ell} \, \eta\,(\eta-2)\, }{24\, (\mathcal{R}+1)^2 }\sum _{a=0}^{\ell+2} r^{a-1}\,\left(\frac{1 + 2 \mathcal{R}}{2 M}\right)^{a-1}\,\mathcal{(-R)}^{-a}\,\sum _{m=a}^{\ell+2} \alpha_{m,i}\,\mathcal{R}^{m}\,{m \choose a} \, ,
\end{split}
\end{equation}
where
\begin{equation}
\begin{split}
\alpha_{m,i}   = \Biggl[ & \theta(m-4)\eta(\eta -2)(\mathcal{R}+1)^2\, {\ell-2 \choose m-4} \frac{  (\ell+3)_{m-4}}{ (5)_{m-4} } \\
   & -\frac{\theta(m-1)\,12 \,\mathcal{R}(2\mathcal{R}+1)\,q_i}{M}\, {\ell+1 \choose m-1}\frac{ (\ell)_{m-1}}{ (2)_{m-1} }
 -24 \mathcal{R}^2\,{\ell+2 \choose m}\frac{ (\ell-1)_m}{ (1)_m } \\ 
   & +\frac{\theta(m-3)\,2 q_i\,\eta  (\mathcal{R}+1)(2\mathcal{R}+1)}{M} \,{\ell-1 \choose m-3}\,\frac{ (\ell+2)_{m-3}}{ (4)_{m-3} } \Biggr] \, .
\end{split}
\end{equation}
Performing the summation over $m$, we  arrive at the final result:
\begin{equation}
\begin{split}
\label{eq:Zirsoln}
Z_i(r)& =
c_{i,\ell}\,\Biggl\{\lambda_i(\ell)\,\left[\frac{24Q^{4}}{\eta}\frac{1}{r}-\frac{6Q^{2}q_{i}}{\eta}+q_{i}r^{2}-r^{3}(\eta-2)\right] 
\\ 
&+\frac{\eta\,(\eta-2)\, }{2\, (\mathcal{R}+1)^2 }\sum _{a>4}^{\ell+2} (-1)^a r^{a-1}\left(\frac{1 + 2 \mathcal{R}}{2 M}\right)^{a-1}\frac{ \Gamma (a+\ell-1)}{\Gamma (a+1) \Gamma (-a+\ell+3)} \times 
\\ 
& \times \Biggl[ \frac{q_i(2 \mathcal{R}+1)}{M}\left(-  \eta \mathcal{R}  \frac{\hypergeom{2}{1}(a-\ell-2,a+\ell-1,a;-\mathcal{R})}{\Gamma (a)}  +   (\mathcal{R}+1)  \frac{\hypergeom{2}{1}(a-\ell-2,a+\ell-1,a-2;-\mathcal{R})}{\Gamma (a-2)} \right) 
\\ 
&+\frac{2 (\mathcal{R}+1)^2  \hypergeom{2}{1}(a-\ell-2,a+\ell-1,a-3;-\mathcal{R})}{\Gamma (a-3)}-\frac{2 \eta  (\eta -2) \mathcal{R}^2 \hypergeom{2}{1}(a-\ell-2,a+\ell-1,a+1;-\mathcal{R})}{\Gamma (a+1)}\Biggr]\Biggr\}
\end{split}
\end{equation}
where
\begin{equation}
\begin{split}
    \lambda_i(\ell)& =\frac{\eta\,(\eta-2)\, (\ell+1)(2 \mathcal{R}+1)^3}{192 M^4 \mathcal{R} (\mathcal{R}+1)^4}\Bigl[ \hypergeom{2}{1}(-\ell,\ell,2;-\mathcal{R}) (q_i (2 \mathcal{R}+1)-2 M (2 \mathcal{R}+3))
    \\
    &\quad -(M (4 (\ell-1) \mathcal{R}-6)+q_i (2 \mathcal{R}+1)) \hypergeom{2}{1}(-\ell-1,\ell,2;-\mathcal{R}) \Bigr] \, .
\end{split}
\end{equation}
We remind  that the above solution has been derived under the assumption that  $\ell>1$, which is the physically interesting case where the gravitational modes are propagating.


\newpage
\renewcommand{\em}{}
\bibliographystyle{utphys}
\addcontentsline{toc}{section}{References}
\bibliography{biblio.bib}

\end{document}